\documentclass[aps,prd,twocolumn,nofootinbib,superscriptaddress]{revtex4-2}

\usepackage[T1]{fontenc}
\usepackage[utf8]{inputenc}
\usepackage{amsmath,amssymb,amsfonts}
\usepackage{graphicx}
\usepackage{booktabs}
\usepackage{hyperref}
\usepackage{bm}
\usepackage{mathtools}
\usepackage{xcolor}
\usepackage{placeins}

\hypersetup{
  colorlinks=true,
  linkcolor=blue,
  citecolor=blue,
  urlcolor=blue
}

\newcommand{\FigPlaceholder}[2]{%
\IfFileExists{#1}{%
  \includegraphics[width=\linewidth]{#1}%
}{%
  \fbox{%
    \parbox[c][4.5cm][c]{0.92\linewidth}{%
      \centering
      Placeholder for #2\\[4pt]
      File \texttt{#1} not found
    }%
  }%
}%
}

\begin{document}

\title{Circular-orbit dynamics and QPO constraints in static Einstein--scalar--Gauss--Bonnet black holes}

\author{Sardor~Murodov}
\email{s.murodov@newuu.uz}
\affiliation{New Uzbekistan University, Movarounnahr Street 1, Tashkent 100000, Uzbekistan}
\affiliation{Tashkent State Technical University, Tashkent 100095, Uzbekistan}

\author{Bekzod~Rahmatov}
\email{rahmatovbekzod@samdu.uz}
\affiliation{University of Tashkent for Applied Sciences, Str. Gavhar 1, Tashkent 100149, Uzbekistan}

\author{Otabek~Umarov}
\email{otabekumarov@kiut.uz}
\affiliation{Department of Exact Sciences, Kimyo International University in Tashkent, Shota Rustaveli Str. 156, Tashkent 100121, Uzbekistan}

\author{Anvar~Kurbaniyazov}
\email{kurbaniyazov85@gmail.com}
\affiliation{Samarkand branch of Tashkent University of Information Technologies named after Muhammad al-Khwarizmi, Ibn Sino street 2A, Samarkand 140110, Uzbekistan}

\author{Faisal~Javed}
\email{faisaljaved.math@gmail.com}
\affiliation{College of Transportation, Tongji University, Shanghai 201804, People's Republic of China}
\affiliation{Research Center of Astrophysics and Cosmology, Khazar University, Baku, AZ1096, 41 Mehseti Street, Azerbaijan}

\author{Javlon~Rayimbaev}
\email{javlon@astrin.uz}
\affiliation{Institute of Theoretical Physics, National University of Uzbekistan, Tashkent 100174, Uzbekistan}

\date{\today}

\begin{abstract}
Einstein--scalar--Gauss--Bonnet (EsGB) gravity provides a physically motivated framework for testing strong-field deviations from the Schwarzschild geometry through scalar hair. We study neutral-particle circular motion and high-frequency quasi-periodic oscillations (HF-QPOs) in static EsGB black holes described by a continued-fraction metric with a single dimensionless deformation parameter \(p\) on the Schwarzschild-connected quadratic-coupling branch. We determine the effective potential, circular-orbit energy and angular momentum, characteristic radii, and orbital and radial epicyclic frequencies, and apply the relativistic precession model to twin-peak QPO data from XTE J1550--564, GRO J1655--40, GRS 1915+105, and M82 X-1. A source-by-source Markov chain Monte Carlo analysis shows that the observed frequency pairs can be reproduced within their uncertainties and that the radial epicyclic frequency carries the main model-level sensitivity to \(p\). However, a controlled prior-sensitivity analysis using uniform and truncated Gaussian priors finds that the marginal posterior of \(p\) closely follows the adopted prior for all four sources. This reflects the intrinsic underconstraint of fitting three correlated parameters \((M,p,r)\) to two measured frequencies. The inferred intervals therefore represent model-dependent compatibility regions rather than an independent measurement or preferred value of the EsGB deformation. The static results provide a baseline for future rotating and multi-observable tests.
\end{abstract}

\maketitle

\section{Introduction}

Black holes are among the most useful natural laboratories for testing gravity, where the field is strongest.  In general relativity, the Schwarzschild and Kerr solutions provide the standard descriptions of isolated nonrotating and rotating black holes, and their geodesic structure underlies much of modern black-hole astrophysics \cite{Einstein1915,Schwarzschild1916,Kerr1963,Carter1968,Bardeen1972,Chandrasekhar1983,Wald1984}.  These solutions have been remarkably successful, but they also motivate a precise question: can the spacetime very close to the horizon contain small, observationally accessible deviations from the predictions of general relativity?  This question is especially timely because compact objects are now probed through several complementary channels, including electromagnetic spectra, gravitational waves, black-hole shadows, and timing observables \cite{Will2014,Berti2015,CardosoPani2019}.

A particularly well-motivated route beyond general relativity is to include higher-curvature corrections.  The Gauss--Bonnet invariant appears naturally in Lovelock gravity and in string-inspired low-energy effective actions, and it becomes dynamically important in four dimensions when it is coupled to an additional scalar degree of freedom \cite{Lovelock1971,BoulwareDeser1985,Zwiebach1985,GrossSloan1987,MetsaevTseytlin1987,NojiriOdintsov2005,NojiriOdintsov2011}.  In this way, EsGB gravity provides a physically transparent framework in which black holes may develop scalar hair while still approaching the usual general-relativistic behaviour far from the source \cite{Kanti1996,AlexeevPomazanov1997,SotiriouZhou2014a,SotiriouZhou2014b,PaniCardoso2009}.  The theory is therefore well suited for studying controlled modifications of the strong-field region without abandoning the successful weak-field limit of general relativity.

Static and rotating EsGB black holes have been investigated in a broad range of contexts.  Previous studies have established the existence of scalarized solutions, the role of the scalar--Gauss--Bonnet coupling, and the domain in which hairy black-hole branches can exist \cite{KleihausKunz2015,KleihausKunz2016,DonevaKanti2018,Silva2018,Antoniou2018,BlazquezSalcedo2018,MinamitsujiIkeda2019}.  Further work has explored stability properties, excited configurations, scalarization thresholds, and the phenomenology of scalarized compact objects \cite{Cunha2019,Collodel2020,Dima2020,Herdeiro2021,EastRipley2021,Witek2019,DonevaYazadjiev2021}.  A practical challenge, however, is that most EsGB black-hole metrics are known numerically rather than as simple closed-form expressions.  For this reason, compact analytical representations are important.  The continued-fraction parametrization provides such a representation by accurately encoding numerical black-hole metrics in terms of a small number of coefficients \cite{RezzollaZhidenko2014,KonoplyaPappasZhidenko2020}.  This approach makes it possible to study geodesics, circular orbits, photon spheres, shadows, and accretion observables without numerically solving the field equations at every step.

Parametrized black-hole metrics play an important role in connecting theory with observations.  They allow one to describe deviations from the Kerr or Schwarzschild geometries in a controlled way and to compute the corresponding changes in lensing, accretion, orbital motion, and quasinormal modes \cite{JohannsenPsaltis2011,Johannsen2013,GlampedakisBabak2006,CardosoPaniRico2014,Bambi2017,Psaltis2008,YunesPretorius2009}.  On the observational side, the Event Horizon Telescope images of M87* and Sgr A* have demonstrated that the near-horizon optical appearance of black holes can be tested directly, while theoretical studies of shadows and strong gravitational lensing clarify which parts of the metric are most relevant for null geodesics \cite{EHT2019I,EHT2019VI,EHT2022SgrA,Vagnozzi2023,PerlickTsupko2022,CunhaHerdeiro2018}.  In the EsGB context, optical and radiative signatures have also been studied through black-hole shadows, thin accretion disks, and quasinormal modes, showing that scalar--Gauss--Bonnet corrections can leave measurable imprints in strong-field observables \cite{HeydariFard2021,Paul2024}. Recently, a first-order eikonal framework has also been used to connect quasinormal modes, black-hole shadows, strong gravitational lensing, and grey-body factors in scalarized black-hole geometries, showing that different wave, optical, and radiative observables can provide complementary probes of strong-field deviations \cite{Lutpi20262}.

The motion of massive particles offers a complementary probe because it depends on the timelike geodesic sector rather than only on photon trajectories.  Circular orbits, the marginally bound orbit, and the innermost stable circular orbit (ISCO) are central to black-hole accretion theory and to the interpretation of emission from the inner disk \cite{NovikovThorne1973,PageThorne1974,ShakuraSunyaev1973,Pringle1981,AbramowiczFragile2013}.  Even a small change in the metric functions can shift the effective potential, modify the binding energy and angular momentum of circular orbits, and alter the radial epicyclic frequency near the ISCO.  Similar strong-field questions have recently been explored in a variety of modified or effective compact-object geometries, including scalar-field, regular, magnetized, nonmetricity-inspired, STVG, Kalb--Ramond, PFDM, wormhole, and other non-Kerr spacetimes \cite{Rahmatov2026Astrophysical,Rahmatov2026Weak,Rahmatov2026Gravitational,Meliyeva2025Theoretical,Rahmatov2025Gravitational,Jumaniyozov2025Black,Saydullayev2025Black,Rahmatov2025QPO,Nishonov2025QPOs,Zulqarnain2026Orbital}. In particular, regular black-hole geometries such as the Dymnikova spacetime have recently been investigated through late-time tails and quasi-resonant modes, emphasizing that perturbative responses may also encode useful information about nonsingular strong-field compact objects \cite{Lutpi20261}.  Related studies of particle acceleration, spinning-particle motion, charged-particle dynamics, current-loop magnetic fields, compact stars, and nonlinear field configurations also show that near-horizon dynamics can be highly sensitive to the underlying gravitational background \cite{Ref2025PDU4801876T,Ref2026EPJC86311R,Ref2025EPJC85953O,Rahmatov2026Magnetic,Banerjee2025Existence,Banerjee2025Effects,Guo2026Analyzing,Guo2026Stability,Khan2026Nonmetric,Ref2026EL15352002A}.

HF-QPOs provide a timing-based route to the same strong-field physics.  These oscillations are observed in the X-ray flux of accreting compact objects and are often associated with characteristic orbital motion in the inner accretion flow \cite{vanDerKlis2006,RemillardMcClintock2006,BelloniStella2014}.  In geodesic models, the observed upper and lower frequencies are related to the azimuthal, radial, and vertical epicyclic frequencies.  In the relativistic precession model, the upper HF-QPO is identified with the orbital frequency and the lower one with the periastron-precession frequency \cite{StellaVietri1998,StellaVietri1999}.  In resonance-based models, simple ratios such as $3:2$ arise from nonlinear coupling between epicyclic modes \cite{AbramowiczKluzniak2001,KluzniakAbramowicz2001,Torok2005,Bursa2004,Rezzolla2003,Kato2001,Stuchlik2013}.  Observationally, twin-peak HF-QPOs have been reported in several important black-hole candidates, including XTE J1550--564, GRO J1655--40, GRS 1915+105, and M82 X-1 \cite{MorganRemillardGreiner1997,Remillard2002,Strohmayer2001,Pasham2014}.

The use of QPOs as tests of gravity is not free from uncertainty, because the physical mechanism that produces HF-QPOs is still debated.  Nevertheless, geodesic frequency models remain valuable because they translate a given spacetime into concrete frequency predictions.  They are therefore useful for identifying which regions of parameter space are compatible with observed frequency pairs and for comparing different gravitational backgrounds.  Recent work has applied QPO and orbital-frequency methods to several non-Kerr or matter-modified spacetimes, showing that timing observables can be combined with mass estimates, shadows, accretion data, and other probes to test strong-field deviations from general relativity \cite{Shermatov2026Circular,Donmez2026Testing,Shermatov2025Phantom,Jumaniyozov2025Radiative,Murodov2025QPOs}.  This motivates a systematic study of HF-QPOs in static EsGB geometries before moving to the more realistic, but also more complicated, rotating case.

In this work we study circular timelike motion and HF-QPO phenomenology in static EsGB black holes.  We focus on the quadratic coupling branch and use the continued-fraction analytical approximation in which the deformation of the metric is controlled by a single dimensionless parameter $p$.  We first examine how $p$ changes the metric functions, the effective potential, the specific energy and angular momentum of circular orbits, and the characteristic radii associated with the horizon, photon sphere, marginally bound orbit, and ISCO.  We then derive the orbital and radial epicyclic frequencies and use the relativistic precession prescription to compare the model with observed twin-peak QPOs from XTE J1550--564, GRO J1655--40, GRS 1915+105, and M82 X-1.  The aim is not to claim a definitive measurement of the EsGB coupling, but to construct a clear static baseline that shows how the scalar--Gauss--Bonnet deformation enters orbital dynamics and timing observables.

The paper is organized as follows. In Section~\ref{sec:esgb_metric} we review the static EsGB metric and the continued-fraction parametrization. Sec~\ref{sec:spacetime_properties} discusses the geometric role of the deformation parameter. In Section~\ref{sec:circular_motion} we study circular timelike motion, including the effective potential, specific energy, specific angular momentum, and characteristic radii. Section~\ref{sec:qpo_section} derives the orbital and epicyclic frequencies and summarizes the QPO prescriptions. Section~\ref{sec:mcmc_constraints} presents the source-by-source MCMC comparison with observed QPO pairs. Section~\ref{sec:discussion} discusses the physical interpretation and limitations of the static analysis, and Section~\ref{sec:conclusion} gives our conclusions.

\section{Static Einstein--scalar--Gauss--Bonnet black holes}
\label{sec:esgb_metric}

We begin with the EsGB action
\begin{equation}
S=\frac{1}{16\pi}\int d^4x \sqrt{-g}\,
\left[
R-\frac{1}{2}\partial_\mu\phi\,\partial^\mu\phi
+\alpha f(\phi)\mathcal{G}
\right],
\label{eq:esgb_action}
\end{equation}
where $R$ is the Ricci scalar, $\phi$ is the scalar field, $\alpha$ is the Gauss--Bonnet coupling constant, and
\begin{equation}
\mathcal{G}
=
R^{2}-4R_{\mu\nu}R^{\mu\nu}
+R_{\mu\nu\rho\sigma}R^{\mu\nu\rho\sigma}
\label{eq:GB_invariant}
\end{equation}
is the Gauss--Bonnet invariant. Static black-hole solutions with scalar hair in EsGB gravity are in general known numerically or perturbatively rather than in simple exact closed form \cite{Antoniou2018,Konoplya2020}. For this reason, it is convenient to work with an analytic parametrization of the numerical solutions.

For a static and spherically symmetric spacetime, we consider the line element
\begin{equation}
ds^2=-N(r)\,dt^2+\frac{B(r)}{N(r)}\,dr^2+r^2(d\theta^2+\sin^2\theta\,d\phi^2),
\label{eq:metric_general}
\end{equation}
where the metric functions $N(r)$ and $B(r)$ describe the deviation from the Schwarzschild geometry. Following Ref.~\cite{Konoplya2020}, we introduce the compact radial coordinate
\begin{equation}
x=1-\frac{r_0}{r},
\label{eq:compact_coordinate}
\end{equation}
where $r_0$ is the event-horizon radius. In terms of $x$, one writes
\begin{equation}
N(x)=x\,A(x),
\label{eq:Nx_def}
\end{equation}
with
\begin{equation}
A(x)=1-\epsilon(1-x)+(a_0-\epsilon)(1-x)^2+\widetilde{A}(x)(1-x)^3,
\label{eq:Ax_def}
\end{equation}
\begin{equation}
B(x)=1+b_0(1-x)+\widetilde{B}(x)(1-x)^2,
\label{eq:Bx_def}
\end{equation}
where
\begin{equation}
\widetilde{A}(x)=
\frac{a_1}{1+\dfrac{a_2 x}{1+\dfrac{a_3 x}{1+\cdots}}},
\qquad
\widetilde{B}(x)=
\frac{b_1}{1+\dfrac{b_2 x}{1+\dfrac{b_3 x}{1+\cdots}}}.
\label{eq:continued_fraction_AB}
\end{equation}
This continued-fraction representation provides an efficient analytic approximation of the numerical EsGB black-hole metrics \cite{Konoplya2020}.

In the present work, we restrict attention to the even-polynomial coupling family and choose the quadratic case
\begin{equation}
f(\phi)=\phi^2,
\label{eq:quadratic_coupling}
\end{equation}
which corresponds to the $n=1$ member of the family considered in Ref.~\cite{Konoplya2020}. To label the family of black-hole solutions, it is convenient to introduce the dimensionless parameter
\begin{equation}
p \equiv
\frac{96\alpha^2 [f'(\phi_0)]^2}{r_0^4},
\label{eq:p_definition}
\end{equation}
where $\phi_0=\phi(r_0)$ is the value of the scalar field at the event horizon. The regularity condition at the horizon requires
\begin{equation}
0\leq p \leq 1 ,
\end{equation}
where $p=0$ corresponds to the Schwarzschild limit, while $p\to 1$ represents the maximal-coupling regime of the corresponding EsGB branch.

Following the continued-fraction parametrization of Ref.~\cite{Konoplya2020}, we use horizon-scaled units in which the radial quantities entering the fitted coefficients are expressed in units of the horizon radius, $r_0=1$. In addition, the fitted analytical expressions used in this work correspond to the normalization $\alpha=1/4$. For the quadratic coupling
\begin{equation}
f(\phi)=\phi^2 ,
\end{equation}
one has $f'(\phi_0)=2\phi_0$, and Eq.~\eqref{eq:p_definition} therefore reduces to
\begin{equation}
p = 24\phi_0^2 .
\end{equation}

The choice $r_0=1$ refers only to the horizon-scaled construction of the fitted continued-fraction coefficients. When physical units are restored, the horizon radius is related to the asymptotic mass through the parameter $\epsilon$ as
\begin{equation}
\epsilon \equiv -\left(1-\frac{2M}{r_0}\right)
= \frac{2M}{r_0}-1 .
\end{equation}
Therefore,
\begin{equation}
\frac{r_0}{M}=\frac{2}{1+\epsilon(p)} .
\end{equation}
In the Schwarzschild limit, $\epsilon=0$ and hence $r_0=2M$.

The $p$-dependent expressions used below are the fitted analytical approximations obtained in Ref.~\cite{Konoplya2020}. They are constructed by solving the EsGB field equations numerically for different values of $p$, matching the resulting metric functions to the continued-fraction representation, and then fitting the extracted coefficients as rational functions of $p$. Thus, the coefficients $\epsilon$, $a_i$, and $b_i$ should be regarded as branch-dependent fitted functions of $p$, not as independent free parameters.

For the quadratic coupling branch considered here, we use the second-order fitted continued-fraction coefficients of Ref.~\cite{Konoplya2020}:
\begin{equation}
\epsilon=\frac{\dfrac{p}{43}-\dfrac{p^2}{201}}{1-\dfrac{105p}{577}},
\label{eq:epsilon_of_p}
\end{equation}
\begin{equation}
a_1=
\frac{\dfrac{63p}{332}-\dfrac{23p^2}{143}}
{-\dfrac{83p^2}{401}+p-\dfrac{223}{259}},
\label{eq:a1_of_p}
\end{equation}
\begin{equation}
a_2=
\frac{-\dfrac{234p^2}{307}+\dfrac{152p}{397}+1}
{\dfrac{73}{221}-\dfrac{73p}{228}},
\label{eq:a2_of_p}
\end{equation}
\begin{equation}
b_1=
\frac{\dfrac{91p}{396}-\dfrac{85p^2}{438}}
{-\dfrac{24p^2}{131}+p-\dfrac{632}{707}},
\label{eq:b1_of_p}
\end{equation}
\begin{equation}
b_2=
\frac{-\dfrac{173p^3}{432}+\dfrac{47p^2}{225}-\dfrac{106p}{203}+1}
{\dfrac{20}{221}-\dfrac{25p}{283}}.
\label{eq:b2_of_p}
\end{equation}
For the EsGB solutions considered here, the asymptotic expansion implies $\beta=\gamma=1$, and therefore
\begin{equation}
a_0=b_0=0.
\label{eq:a0b0_zero}
\end{equation}

In this way, the metric is no longer described by an arbitrary set of fitting coefficients, but by a smaller and more physical parameter set. Throughout this work we therefore regard
\begin{equation}
\epsilon=\epsilon(p),\qquad
a_i=a_i(p),\qquad
b_i=b_i(p),
\label{eq:coeffs_of_p}
\end{equation}
so that the relevant phenomenological parameters are reduced to the smaller set $(M,p,r)$, where $M$ is the black-hole mass, $p$ is the effective EsGB deformation parameter, and $r$ is the orbital radius entering the geodesic analysis.

The Schwarzschild limit is recovered for
\begin{equation}
p=0,\qquad r_0=2M,\qquad \epsilon=0,\qquad a_i=b_i=0,
\label{eq:schwarzschild_limit_params}
\end{equation}
which implies
\begin{equation}
N(r)=1-\frac{2M}{r},
\qquad
B(r)=1.
\label{eq:schwarzschild_metric_functions}
\end{equation}
Thus Eq.~\eqref{eq:metric_general} reduces to the familiar Schwarzschild metric,
\begin{equation}
ds^2
=
-\left(1-\frac{2M}{r}\right)dt^2
+\frac{dr^2}{1-\frac{2M}{r}}
+r^2(d\theta^2+\sin^2\theta\,d\phi^2).
\label{eq:schwarzschild_metric}
\end{equation}

\subsection{Physical meaning of the parametrization coefficients}
\label{subsec:param_interpretation}

The coefficients $\epsilon$, $a_0$, $b_0$, and $a_i,b_i$ $(i\geq1)$ should not be viewed as independent physical observables. Rather, they provide an analytic representation of a specific static EsGB black-hole solution. The compact coordinate $x$ separates the horizon region ($x=0$) from the asymptotic region ($x=1$), while the continued-fraction structure ensures rapid convergence of the approximation in the domain outside the event horizon \cite{Konoplya2020}.

Within this parametrization, the parameter $\epsilon$ measures the leading global deviation from the Schwarzschild relation between the horizon radius and the asymptotic mass. The coefficients $a_0$ and $b_0$ control the first asymptotic corrections and vanish for the present EsGB family, while the higher-order coefficients $a_i$ and $b_i$ encode the detailed strong-field structure near the horizon. In practice, these coefficients are obtained by matching the parametrized metric to the corresponding numerical EsGB solution \cite{Konoplya2020}.

For the quadratic coupling studied here, it is therefore more transparent to regard the spacetime as controlled by the effective deformation parameter $p$, with the coefficients $\epsilon$, $a_i$, and $b_i$ treated as branch-dependent derived quantities. This choice is both physically cleaner and numerically more stable than attempting to vary all continued-fraction coefficients independently.

\begin{figure*}[ht!]
    \centering
    \includegraphics[width=0.45\linewidth]{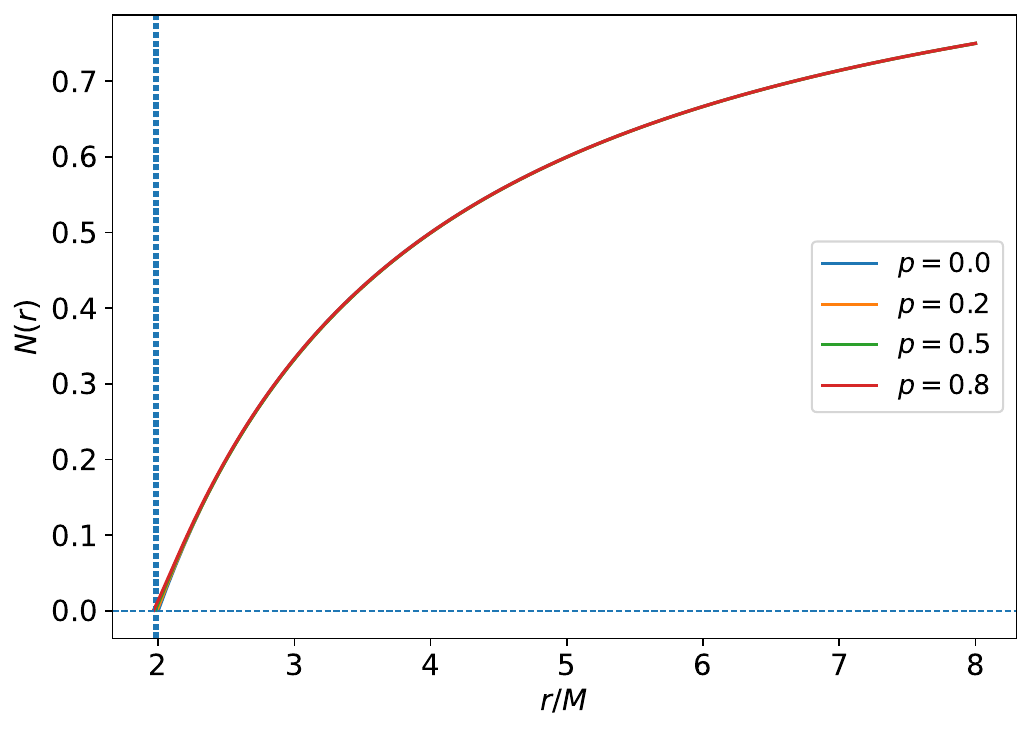}
    \includegraphics[width=0.45\linewidth]{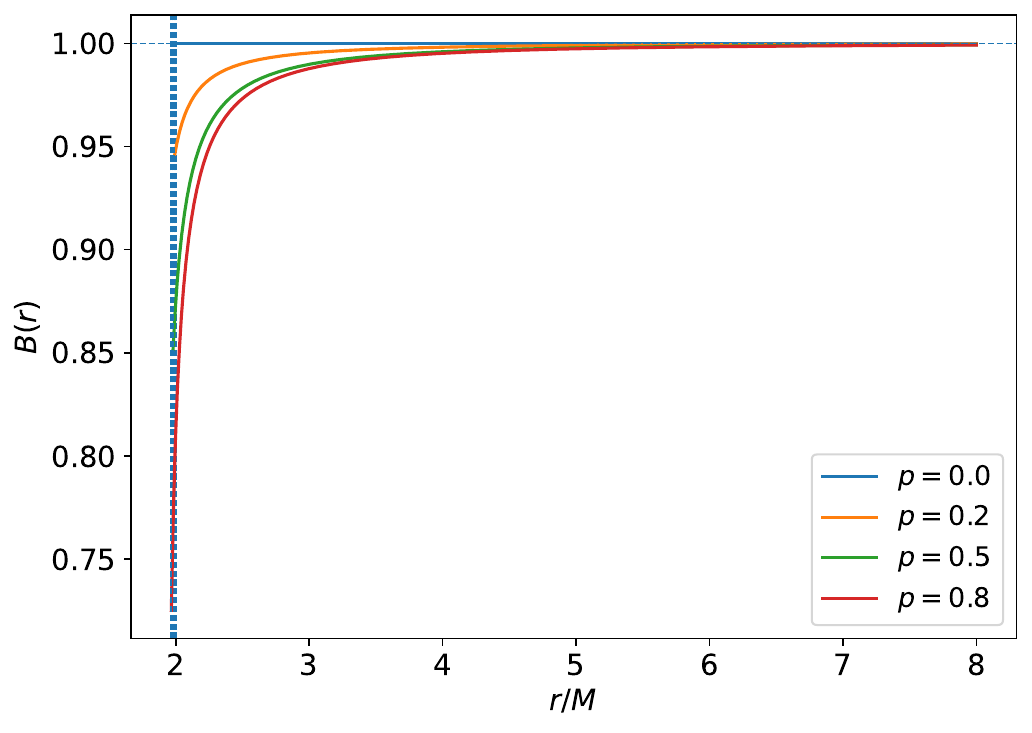}
    \includegraphics[width=0.45\linewidth]{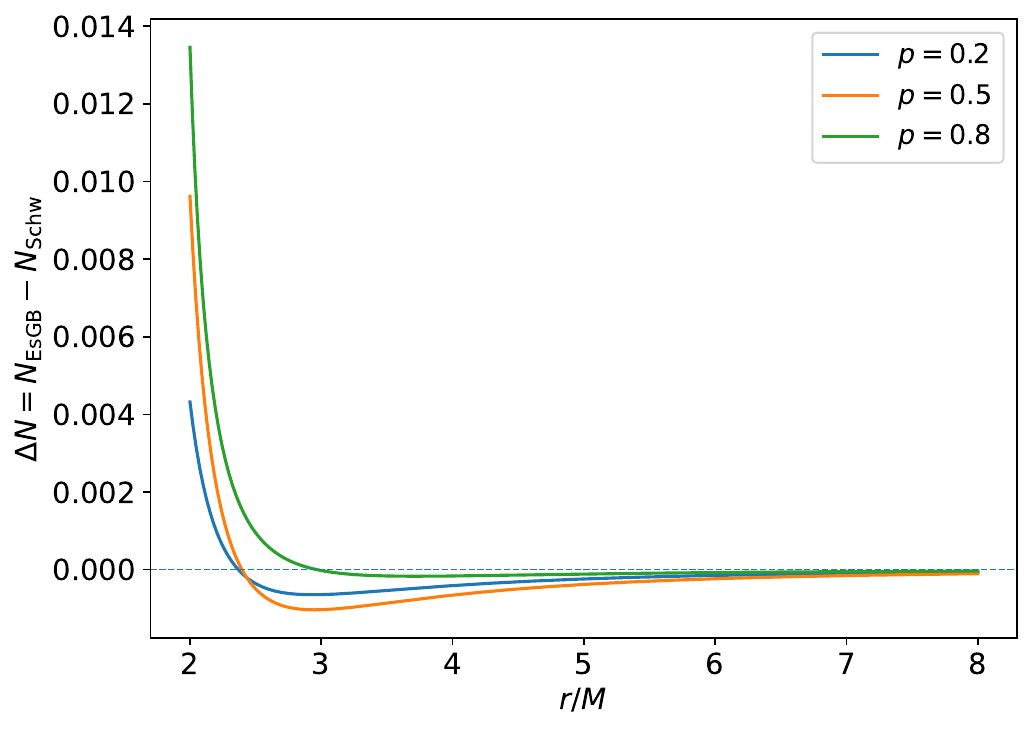}
    \includegraphics[width=0.45\linewidth]{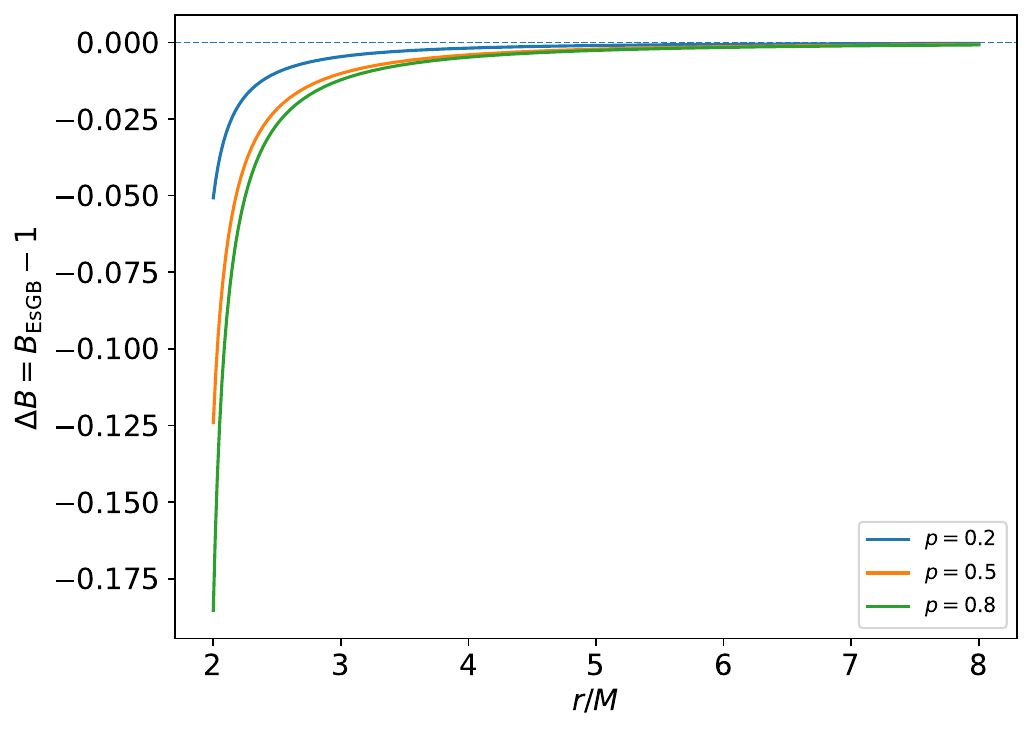}
    \caption{Radial behaviour of the metric functions in the EsGB black hole spacetime and their deviations from the Schwarzschild geometry. The upper panels show the functions $N(r)$ and $B(r)$ for different values of the deformation parameter $p$, while the lower panels present the corresponding deviations $\Delta N=N_{\rm EsGB}-N_{\rm Schw}$ and $\Delta B=B_{\rm EsGB}-1$. The deviations are more pronounced in the near-horizon region and rapidly decrease at larger radii, indicating that the EsGB correction mainly affects the strong-field domain of the spacetime.}
    \label{fig:metric}
\end{figure*}

Figure~\ref{fig:metric} shows how the EsGB deformation changes the background geometry relative to the Schwarzschild black hole. The function $N(r)$ describes the temporal component of the metric and is closely related to the gravitational redshift and the effective gravitational potential, whereas $B(r)$ controls the radial part of the geometry and determines how radial distances are measured in the curved spacetime. Therefore, changes in these two functions directly reflect how the spacetime structure is modified by the EsGB parameter $p$.

The most important physical feature seen in this figure is that the deviations from the Schwarzschild case are localized mainly near the event horizon. This is expected because the Gauss--Bonnet contribution is curvature-dependent and becomes dynamically more relevant in regions where the gravitational field is strong. As the radial distance increases, the curvature weakens and both $\Delta N$ and $\Delta B$ quickly approach zero. This means that the EsGB spacetime smoothly recovers the Schwarzschild behaviour in the weak-field region.

This result is important for the later analysis of particle motion and QPO frequencies. Since circular orbits, the ISCO position, and epicyclic frequencies are highly sensitive to the near-horizon geometry, even small deviations in $N(r)$ and $B(r)$ can lead to measurable changes in the orbital dynamics. At the same time, the rapid decay of the deviations at large radii shows that the parameter $p$ does not strongly modify the outer spacetime, but mainly introduces corrections in the strong-gravity region where observational signatures of modified gravity are expected to appear.

\section{Spacetime properties}
\label{sec:spacetime_properties}

Having fixed the quadratic Einstein--scalar--Gauss--Bonnet coupling and the Schwarzschild-connected branch, we now summarize the basic geometric properties of the corresponding static black holes. In the present parametrization, the spacetime is labeled by the dimensionless deformation parameter $p$, with
\begin{equation}
p=0
\end{equation}
corresponding to the Schwarzschild limit and increasing values of $p$ describing progressively stronger scalar--Gauss--Bonnet deformations \cite{Konoplya2020}.

The event horizon is located at
\begin{equation}
r=r_0,
\end{equation}
or equivalently at $x=0$ in the compact radial coordinate
\begin{equation}
x=1-\frac{r_0}{r}.
\end{equation}
By construction, the metric function $N(r)$ vanishes at the horizon,
\begin{equation}
N(r_0)=0,
\end{equation}
while the exterior region corresponds to $r>r_0$ or $0<x<1$. Spatial infinity is reached in the limit $x\to1$.

The solutions considered here are asymptotically flat. Accordingly, at large distances the metric functions approach their Schwarzschild form,
\begin{equation}
N(r)\to 1-\frac{2M}{r}+\mathcal{O}(r^{-2}),
\qquad
B(r)\to 1+\mathcal{O}(r^{-1}),
\qquad r\to\infty,
\label{eq:asymptotic_flatness_sec3}
\end{equation}
which ensures that the strong-field EsGB corrections remain localized in the near-horizon region \cite{Antoniou2018,Konoplya2020}. In the present quadratic case, the geometry is therefore most naturally viewed as a one-parameter deformation of Schwarzschild controlled by $p$.

A convenient way to illustrate the effect of the deformation is to consider a small set of representative values,
\begin{equation}
p=\{0,\;0.2,\;0.5,\;0.8\},
\label{eq:representative_p_values}
\end{equation}
which correspond, respectively, to the Schwarzschild limit, weak deformation, intermediate deformation, and a strong but still nonextremal EsGB deformation. For each value of $p$, the coefficients $\epsilon$, $a_i$, and $b_i$ are uniquely determined through the fitted relations introduced in Section~\ref{sec:esgb_metric}, and the full metric is therefore fixed once $(M,p)$ is specified.

The optical sector already indicates that the parameter $p$ has a systematic effect on the strong-field geometry. In the static EsGB solutions considered here, the photon-sphere radius remains close to its Schwarzschild value and shows only a weak, nonmonotonic dependence on $p$. This indicates that the null circular-orbit sector is affected by the scalar--Gauss--Bonnet deformation, but the effect remains small for the representative parameter values used in this work. Although these observables are not the primary focus of the present work, they provide useful evidence that the deformation parameter modifies the near-horizon structure in a physically meaningful way.

At the same time, the regime $p\to1$ should be treated with care. In the analytic approximation of Ref.~\cite{Konoplya2020}, the continued-fraction expansion converges more slowly close to the maximal-coupling limit, and the same study points out that very large couplings are likely associated with unstable configurations. For this reason, the phenomenological analysis developed below will focus on nonextremal values of $p$, where the parametrized metric remains both accurate and physically better motivated.

\begin{table}[t]
\caption{Characteristic radii of the EsGB black hole for selected values of $p$.}
\label{tab:characteristic_radii_p}
\centering
\renewcommand{\arraystretch}{1.12}
\begin{tabular}{ccccc}
\toprule
$p$ & $r_h/M$ & $r_{\rm ph}/M$ & $r_{\rm MBO}/M$ & $r_{\rm ISCO}/M$ \\
\midrule
0.0 & 2.000 & 3.000 & 4.000 & 6.000 \\
0.2 & 1.991 & 3.002 & 4.007 & 6.013 \\
0.5 & 1.977 & 3.004 & 4.012 & 6.021 \\
0.8 & 1.965 & 2.997 & 4.001 & 6.005 \\
\bottomrule
\end{tabular}
\end{table}

Table~\ref{tab:characteristic_radii_p} summarizes the characteristic radii of the EsGB black hole for several representative values of the deformation parameter $p$. At $p=0$, the spacetime reduces to the Schwarzschild case, and the expected values $r_h/M=2$, $r_{\rm ph}/M=3$, $r_{\rm MBO}/M=4$, and $r_{\rm ISCO}/M=6$ are recovered. This provides a useful consistency check for the numerical calculation.

As the parameter $p$ increases, the horizon radius slightly decreases, indicating that the EsGB deformation makes the effective black-hole boundary marginally smaller in these units. The photon-sphere, marginally bound orbit, and ISCO radii remain close to their Schwarzschild values, but they show small nonmonotonic shifts. This behaviour means that the scalar--Gauss--Bonnet correction does not simply rescale all orbital radii in one direction. Instead, it modifies the near-horizon effective geometry in a more subtle way.

Physically, these small shifts are important because each radius controls a different aspect of black-hole dynamics. The photon sphere is related to light propagation and shadow formation, the marginally bound orbit separates bound from unbound timelike circular motion, and the ISCO determines the inner edge of stable circular orbits. Therefore, even modest changes in these radii can influence accretion dynamics, epicyclic frequencies, and QPO-related observables in the strong-field region.

The main conclusion of this section is therefore simple: once the quadratic coupling family and the Schwarzschild-connected branch are fixed, the static EsGB black-hole geometry may be treated as a controlled one-parameter deformation of Schwarzschild, with the parameter $p$ governing the strength of the deviation. This makes the subsequent analysis of circular motion and geodesic HF-QPOs both transparent and numerically manageable.

\section{Circular motion of test particles}
\label{sec:circular_motion}

We now turn to timelike geodesic motion in the static EsGB spacetime introduced in Section~\ref{sec:esgb_metric}. Since the metric functions $N(r)$ and $B(r)$ are uniquely fixed once the black-hole mass $M$ and the effective EsGB deformation parameter $p$ are specified, all orbital quantities derived below should be regarded as branch-dependent functions of $(r,p)$ \cite{Konoplya2020}. In this sense, circular motion provides the first direct link between the spacetime deformation and the observable timing sector.

\subsection{Effective potential}

We restrict attention to equatorial motion,
\begin{equation}
\theta=\frac{\pi}{2},
\end{equation}
without loss of generality. The Lagrangian associated with the line element \eqref{eq:metric_general} 
takes the form
\begin{equation}
2\mathcal{L} = -N(r)\dot{t}^{2}
+\frac{B(r)}{N(r)}\dot{r}^{2}
+r^{2}\dot{\phi}^{2}.
\end{equation}
where an overdot denotes differentiation with respect to the proper time $\tau$.

Because the spacetime is stationary and spherically symmetric, the specific energy $E$ and specific angular momentum $L$ are conserved:
\begin{equation}
E=N(r)\dot{t},
\qquad
L=r^2\dot{\phi}.
\label{eq:conserved_quantities_p}
\end{equation}
Using the timelike normalization condition
\begin{equation}
u^\mu u_\mu=-1,
\end{equation}
one finds the radial equation
\begin{equation}
\dot{r}^{\,2}
=
\frac{E^2-V_{\rm eff}(r)}{B(r)},
\label{eq:radial_equation_p}
\end{equation}
with the effective potential
\begin{equation}
V_{\rm eff}(r)=N(r)\left(1+\frac{L^2}{r^2}\right).
\label{eq:effective_potential_p}
\end{equation}
The dependence on the EsGB deformation parameter $p$ is implicit through the metric function $N(r)$.

Equation \eqref{eq:effective_potential_p} shows that the scalar--Gauss--Bonnet deformation influences the motion of massive particles primarily through the modification of the redshift function $N(r)$. Therefore, once $p$ departs from zero, the entire profile of the effective potential shifts relative to the Schwarzschild case, leading to changes in the location and stability of circular orbits.


For the EsGB branch considered in this work, $a_0=0$.
Thus Eq.~\eqref{eq:effective_potential_p} reduces to

\begin{equation}
\begin{aligned}
V_{\rm eff}(r)
&=
\left(1-\frac{r_0}{r}\right)
\left[
1-\epsilon\frac{r_0}{r}
-\epsilon\left(\frac{r_0}{r}\right)^2
+\frac{a_1\left(\frac{r_0}{r}\right)^3}
{1+a_2\left(1-\frac{r_0}{r}\right)}
\right] \\
&\quad \times
\left(1+\frac{L^2}{r^2}\right).
\end{aligned}
\label{eq:Veff_EsGB_coeff}
\end{equation}

\begin{figure}[ht!]
    \centering
    \includegraphics[width=0.9\linewidth]{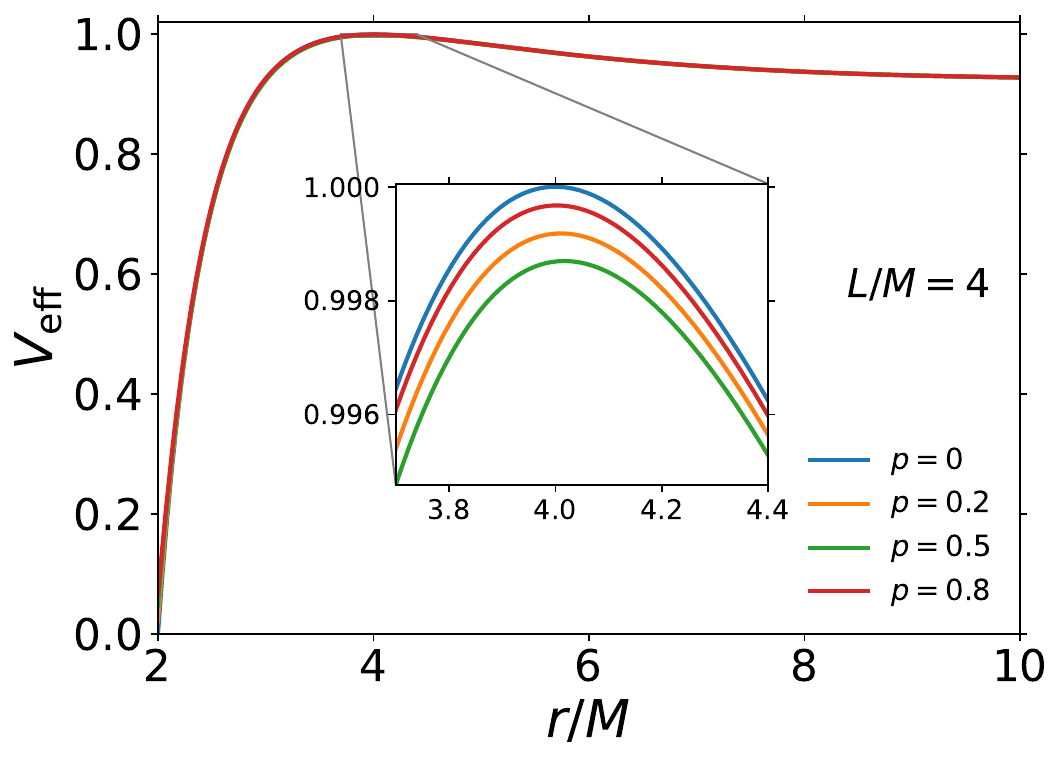}
    \caption{Effective potential for timelike equatorial motion around the EsGB black hole for a fixed angular momentum $L/M=4$. Different curves correspond to different values of the deformation parameter $p$. The variation of $p$ slightly modifies both the depth and the radial position of the potential well, showing that the EsGB correction shifts the location and stability properties of circular orbits with respect to the Schwarzschild limit.}
    \label{fig:Veff}
\end{figure}

Figure~\ref{fig:Veff} shows the effective potential governing the radial motion of massive test particles in the equatorial plane. In this representation, circular orbits are associated with extrema of $V_{\rm eff}$, while stable circular orbits correspond to local minima of the potential. Therefore, the shape of the potential well provides direct information about where particles can remain on bounded circular trajectories and how stable such motion is against small radial perturbations.

The figure demonstrates that the EsGB deformation parameter $p$ produces a visible, although moderate, change in the effective potential. As $p$ increases, the minimum of the potential well is slightly shifted and its depth is modified. Physically, this means that the scalar--Gauss--Bonnet correction changes the balance between the attractive gravitational field and the centrifugal contribution caused by the particle angular momentum. Since this balance determines the position of circular orbits, the deformation parameter can move the preferred orbital radius inward or outward relative to the Schwarzschild case.

This result is important for the QPO analysis, because the characteristic frequencies of particle motion are determined by the local curvature of the effective potential near circular orbits. A change in the depth or curvature of the potential well leads to a corresponding change in the radial epicyclic frequency and in the location of the innermost stable circular orbit. Hence, even small deviations in $V_{\rm eff}$ may leave observable signatures in the high-frequency QPO spectrum, especially in the strong-field region close to the black hole.

\subsection{Circular orbits}

Circular timelike orbits at radius $r=r_c$ satisfy
\begin{equation}
\dot r=0,
\qquad
\frac{dV_{\rm eff}}{dr}\bigg|_{r=r_c}=0.
\label{eq:circular_conditions_p}
\end{equation}
These conditions imply
\begin{equation}
V_{\rm eff}(r_c)=E^2,
\end{equation}
together with the extremum condition for the potential. Solving for the conserved quantities yields the specific energy and specific angular momentum of the circular orbit,
\begin{equation}
E_c^2(r_c)
=
\frac{2N(r_c)^2}{2N(r_c)-r_cN'(r_c)},
\label{eq:energy_circular_p}
\end{equation}
\begin{equation}
L_c^2(r_c)
=
\frac{r_c^3N'(r_c)}{2N(r_c)-r_cN'(r_c)}.
\label{eq:angmom_circular_p}
\end{equation}

The existence of circular timelike motion requires
\begin{equation}
2N(r_c)-r_cN'(r_c)>0,
\label{eq:circular_existence_condition_p}
\end{equation}
together with $L_c^2>0$. In practice, both conditions are satisfied in the physically relevant region outside the horizon for the Schwarzschild-connected branch, but the allowed orbital domain and the detailed radial profiles of $E_c$ and $L_c$ depend on the chosen value of $p$.

\begin{figure*}[ht!]
    \centering
    \includegraphics[width=0.45\linewidth]{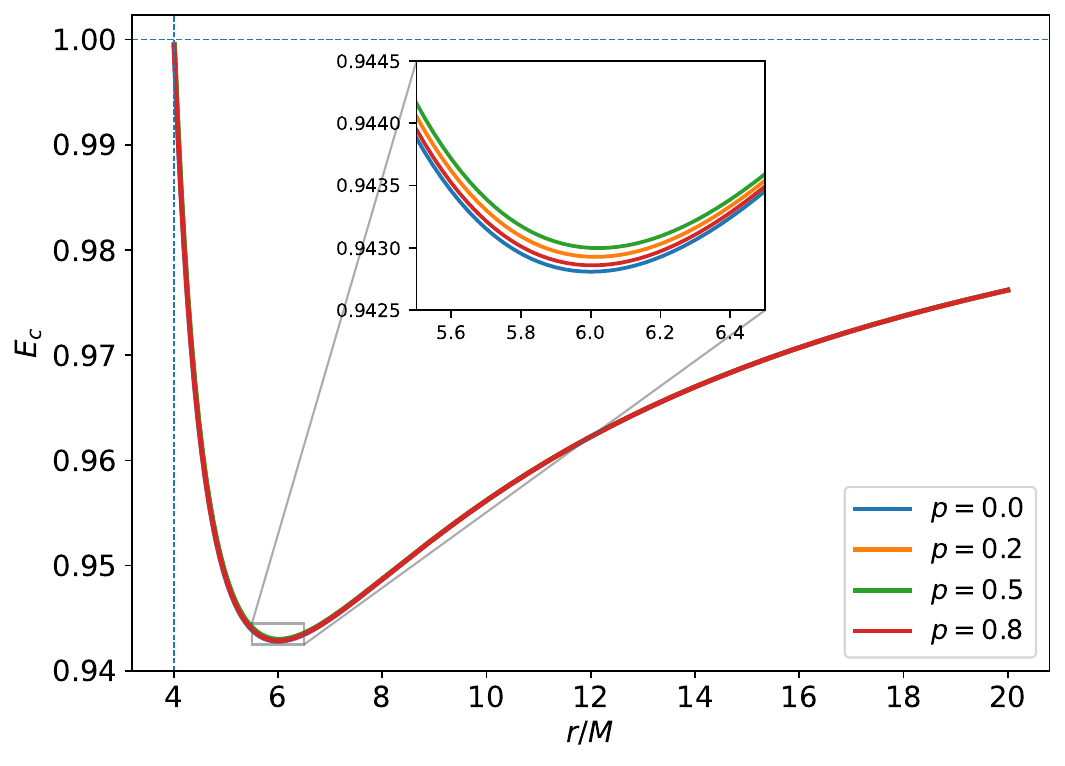}
    \includegraphics[width=0.45\linewidth]{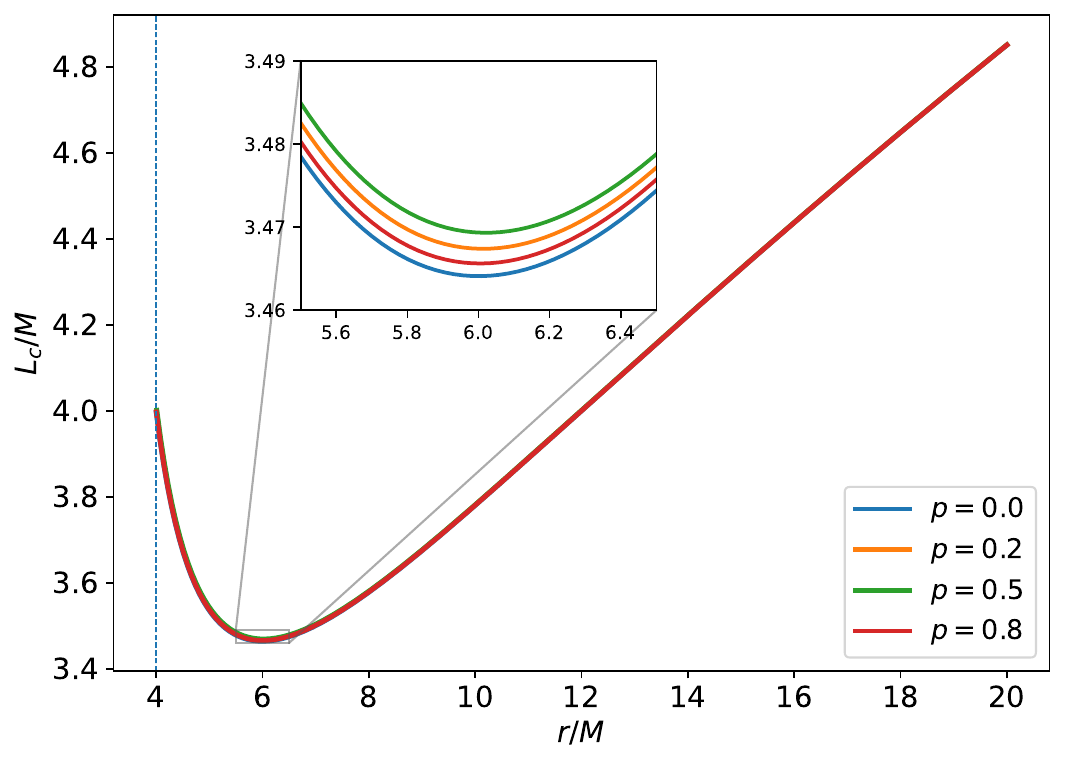}
    \includegraphics[width=0.45\linewidth]{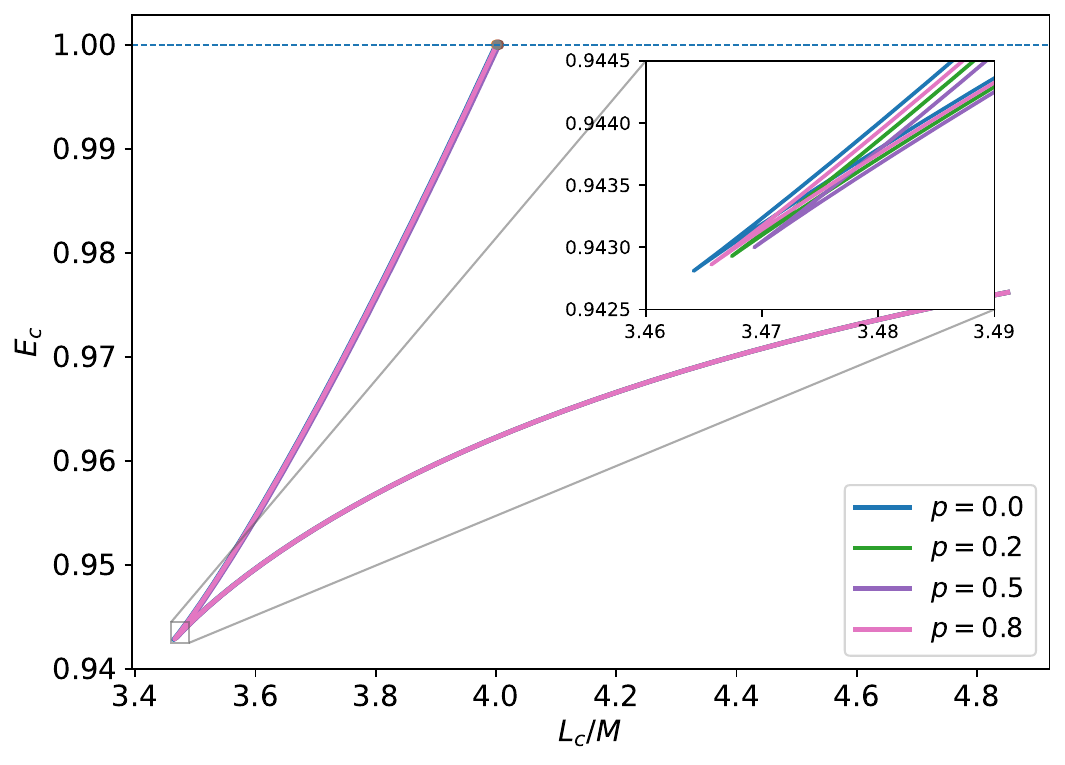}
    \caption{Specific energy $E_c$, specific angular momentum $L_c/M$, and the parametric $E_c$--$L_c$ relation for timelike circular orbits around the EsGB black hole. Different curves correspond to different values of the deformation parameter $p$. The minimum of the angular-momentum profile indicates the onset of marginal stability, while the condition $E_c=1$ separates bound and unbound circular motion. The visible separation between the curves shows that the scalar--Gauss--Bonnet correction modifies the energetic and angular-momentum requirements for circular orbits.}
    \label{fig:specificEL}
\end{figure*}

Figure~\ref{fig:specificEL} shows how the EsGB deformation affects the energy and angular momentum of massive particles moving along circular timelike orbits. The specific energy $E_c$ represents the energy per unit rest mass required for a particle to remain on a circular orbit at a given radius. Values with $E_c<1$ correspond to bound orbits, whereas the condition $E_c=1$ marks the marginally bound orbit. Therefore, changes in the $E_c(r)$ profile indicate how the EsGB parameter modifies the binding energy of particles in the strong gravitational field.

The behaviour of the specific angular momentum $L_c/M$ gives complementary information about orbital stability. For circular motion, the particle angular momentum must balance the gravitational attraction of the black hole. The minimum of the $L_c(r)$ curve is especially important because it is associated with the transition from stable to marginally stable circular motion. In this sense, the location of this minimum provides a useful indicator of the ISCO position. A shift of the minimum with increasing $p$ means that the scalar--Gauss--Bonnet correction changes the radius at which stable circular motion can exist.

The parametric $E_c$--$L_c$ panel combines these two quantities and shows how the energy and angular-momentum requirements are correlated for circular orbits. The separation between the curves for different values of $p$ demonstrates that the EsGB deformation does not merely rescale the orbital parameters, but changes the structure of the circular-orbit family itself. This is physically important because both $E_c$ and $L_c$ enter the description of accretion dynamics, ISCO properties, and epicyclic frequencies. Hence, the figure confirms that the parameter $p$ can leave observable imprints on the motion of matter in the inner region of the accretion flow, where high-frequency QPOs are expected to originate.

\subsection{Marginally stable and marginally bound orbits}

The ISCO is determined by the marginal stability condition
\begin{equation}
\frac{d^2V_{\rm eff}}{dr^2}\bigg|_{r=r_{\rm ISCO}}=0,
\qquad
L=L_c(r_{\rm ISCO},p).
\label{eq:isco_condition_p}
\end{equation}
The ISCO marks the inner edge of stable circular motion and therefore plays a central role in both accretion physics and geodesic timing models. Since $N(r)$ depends on $p$, the ISCO radius also becomes a branch-dependent quantity,
\begin{equation}
r_{\rm ISCO}=r_{\rm ISCO}(p).
\end{equation}

A second characteristic radius is the marginally bound orbit (MBO), defined by the condition
\begin{equation}
E_c(r_{\rm MBO},p)=1.
\label{eq:mbo_condition_p}
\end{equation}
This radius separates bound and unbound circular motion. As in the case of the ISCO, the MBO shifts away from its Schwarzschild value once the scalar--Gauss--Bonnet deformation is turned on.

For reference, in the Schwarzschild limit one recovers
\begin{equation}
r_{\rm ISCO}=6M,
\qquad
r_{\rm MBO}=4M.
\label{eq:schwarzschild_radii_timelike}
\end{equation}
Accordingly, the deviation of $r_{\rm ISCO}(p)$ and $r_{\rm MBO}(p)$ from these values provides a simple quantitative measure of how strongly the EsGB geometry modifies circular timelike motion.

\begin{figure*}[ht!]
    \centering
    \includegraphics[width=0.45\linewidth]{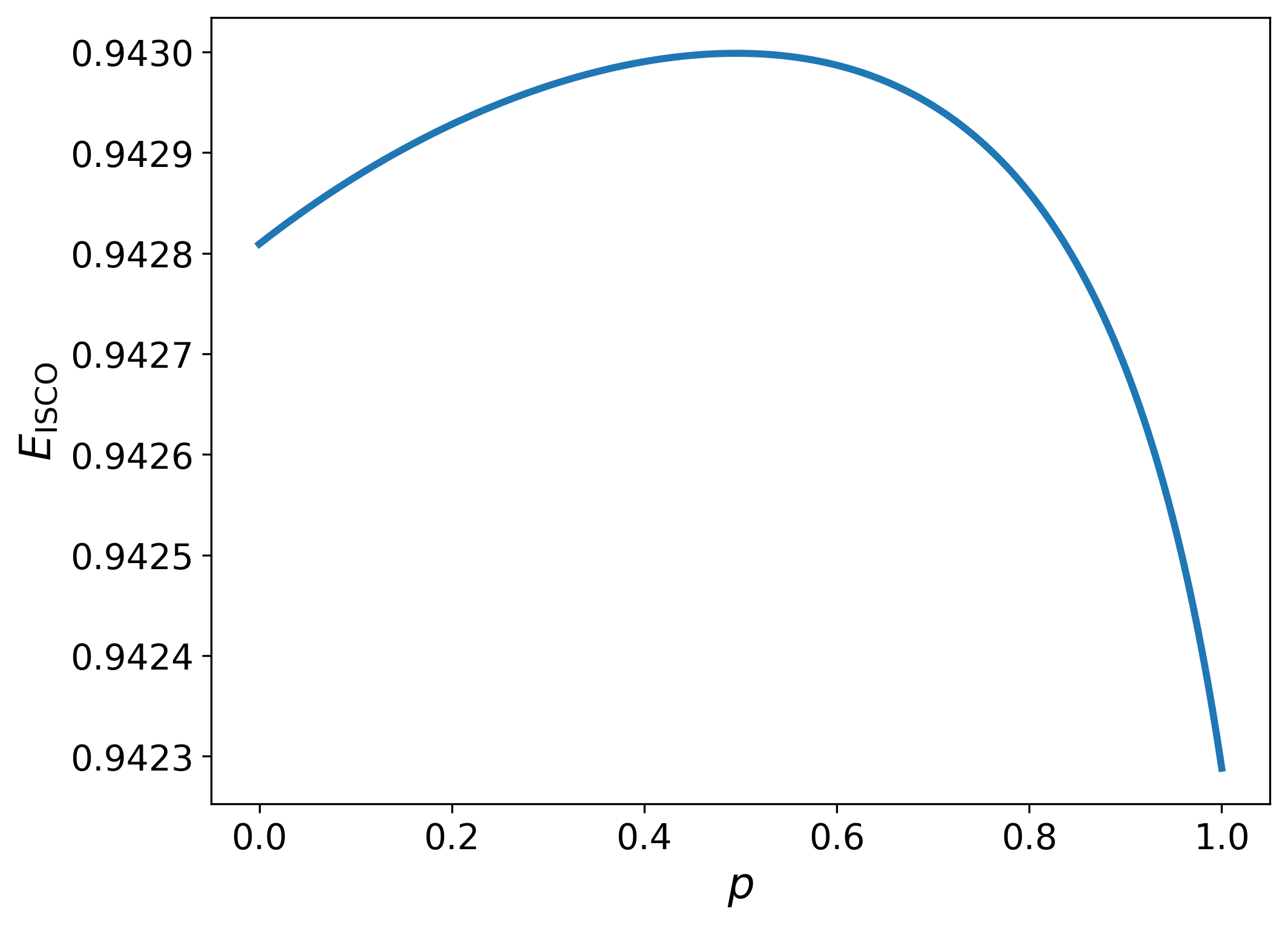}
    \includegraphics[width=0.45\linewidth]{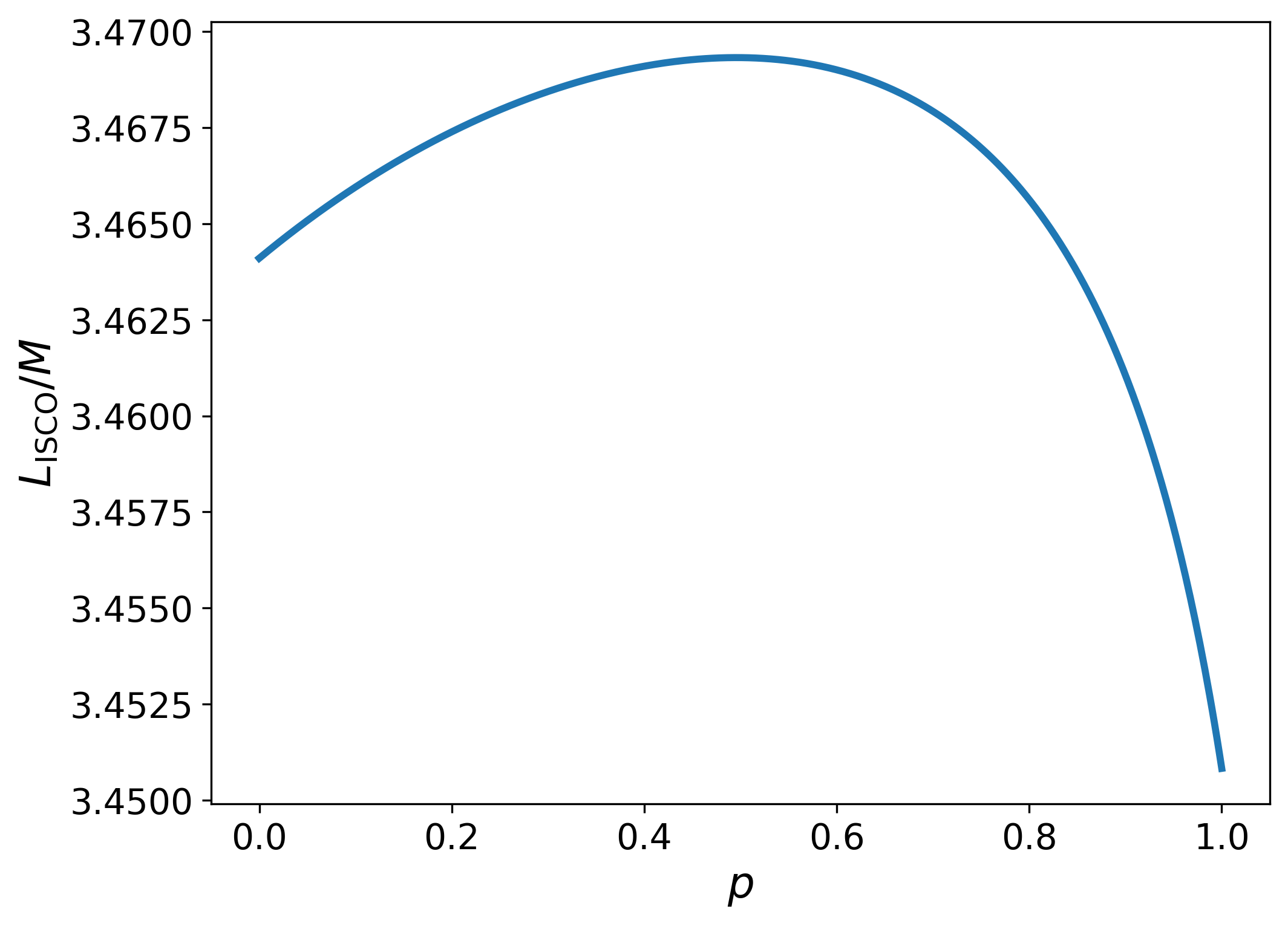}
    \caption{Dependence of the specific energy $E_{\rm ISCO}$ and specific angular momentum $L_{\rm ISCO}/M$ at the innermost stable circular orbit on the EsGB deformation parameter $p$. The left panel shows the ISCO energy, while the right panel shows the corresponding angular momentum. At $p=0$, both quantities recover their Schwarzschild values. The nonmonotonic behaviour indicates that the EsGB deformation changes the near-horizon effective potential in a nontrivial way, rather than simply shifting the orbital radius.}
    \label{fig:ISCOEJ}
\end{figure*}

Figure~\ref{fig:ISCOEJ} shows how the energetic properties of the innermost stable circular orbit respond to the EsGB deformation parameter $p$. The ISCO is a particularly important orbit because it marks the inner boundary of stable circular motion. Inside this radius, small radial perturbations grow and a massive particle can no longer maintain a stable circular trajectory. Therefore, the quantities $E_{\rm ISCO}$ and $L_{\rm ISCO}/M$ provide direct information about the binding energy and angular-momentum requirement of matter moving in the innermost part of the accretion flow.

The left panel demonstrates that $E_{\rm ISCO}$ changes only mildly with $p$, but its behaviour is not purely monotonic. Starting from the Schwarzschild value at $p=0$, the ISCO energy first increases slightly and then decreases for larger values of the deformation parameter. Physically, this means that the scalar--Gauss--Bonnet correction first makes the marginally stable orbit slightly less bound, but at stronger deformation it increases the binding of matter at the ISCO. Such behaviour reflects the fact that the deformation modifies the local shape of the effective potential near the black hole.

A similar trend is observed for $L_{\rm ISCO}/M$ in the right panel. The angular momentum required to maintain a marginally stable circular orbit initially increases with $p$ and then decreases when the deformation becomes larger. This shows that the EsGB correction changes the balance between gravitational attraction and centrifugal support. Since this balance determines the existence and stability of circular orbits, the nonmonotonic variation of $L_{\rm ISCO}/M$ is a clear signature that the near-horizon orbital structure is modified in a nontrivial manner.

These results are important for the QPO analysis. The ISCO determines the innermost region where stable orbital motion can occur, and the characteristic frequencies of particle oscillations are highly sensitive to the geometry in this region. Hence, even small variations in $E_{\rm ISCO}$ and $L_{\rm ISCO}/M$ can influence the epicyclic frequencies and the location where high-frequency QPOs are generated. The figure therefore shows that the parameter $p$ may leave observable imprints through the dynamics of matter close to the black hole.

\subsection{Photon sphere}

Although the primary focus of the present section is timelike motion, it is useful to record the location of the photon sphere, since it also reflects the structure of the strong-field region. For the metric under consideration, the photon-sphere radius is determined by
\begin{equation}
2N(r_{\rm ph})-r_{\rm ph}N'(r_{\rm ph})=0.
\label{eq:photon_sphere_condition_p}
\end{equation}
In the Schwarzschild case,
\begin{equation}
r_{\rm ph}=3M,
\end{equation}
whereas for static EsGB black holes the photon-sphere radius becomes a deformation-dependent quantity,
\begin{equation}
r_{\rm ph}=r_{\rm ph}(p).
\end{equation}
For the representative branch considered in this work, the photon-sphere radius remains very close to the Schwarzschild value and exhibits a weak nonmonotonic dependence on $p$. This behaviour shows that the EsGB correction modifies the null circular-orbit sector without producing a large global displacement of the photon sphere.

\subsection{Physical interpretation}

The results above show that the geodesic structure of the static EsGB spacetime is fully controlled by the pair $(M,p)$. Once a specific value of $p$ is chosen, the effective potential, the energetics of circular motion, and the characteristic radii of the spacetime are all fixed. This makes the deformation parameter $p$ the natural quantity through which the static EsGB geometry enters the timing analysis.

From the phenomenological point of view, the most important outcome is that the scalar--Gauss--Bonnet sector modifies the shape of the effective potential and, as a consequence, shifts the ISCO, the MBO, and the photon sphere relative to their Schwarzschild values. These shifts propagate directly into the orbital and epicyclic frequencies, which form the basis of geodesic HF-QPO models. The next section is therefore devoted to the derivation of those frequencies and to the construction of the corresponding QPO combinations.

\begin{figure*}[ht!]
    \centering
    \includegraphics[width=0.45\linewidth]{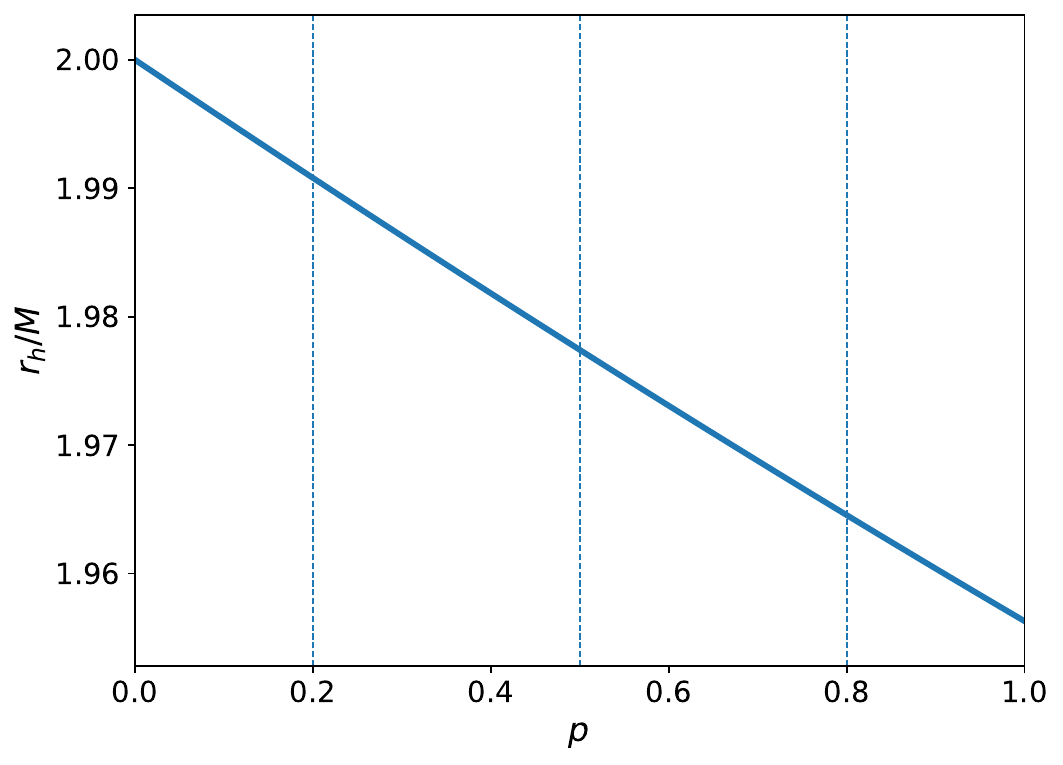}
    \includegraphics[width=0.45\linewidth]{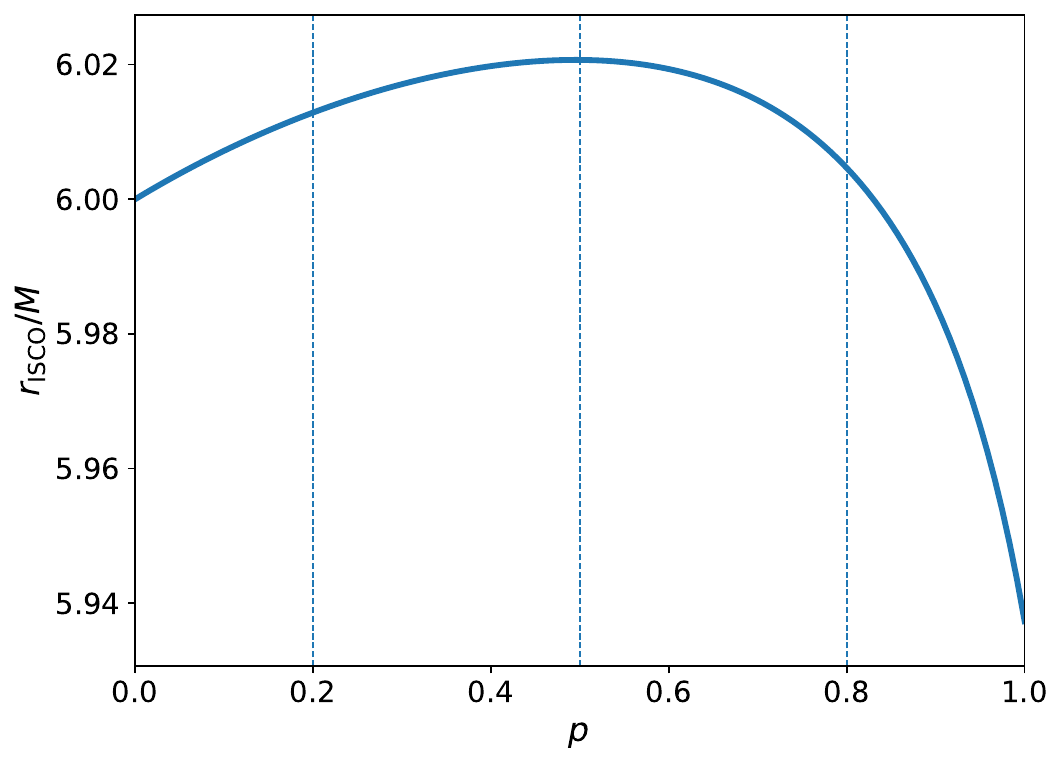}
    \includegraphics[width=0.45\linewidth]{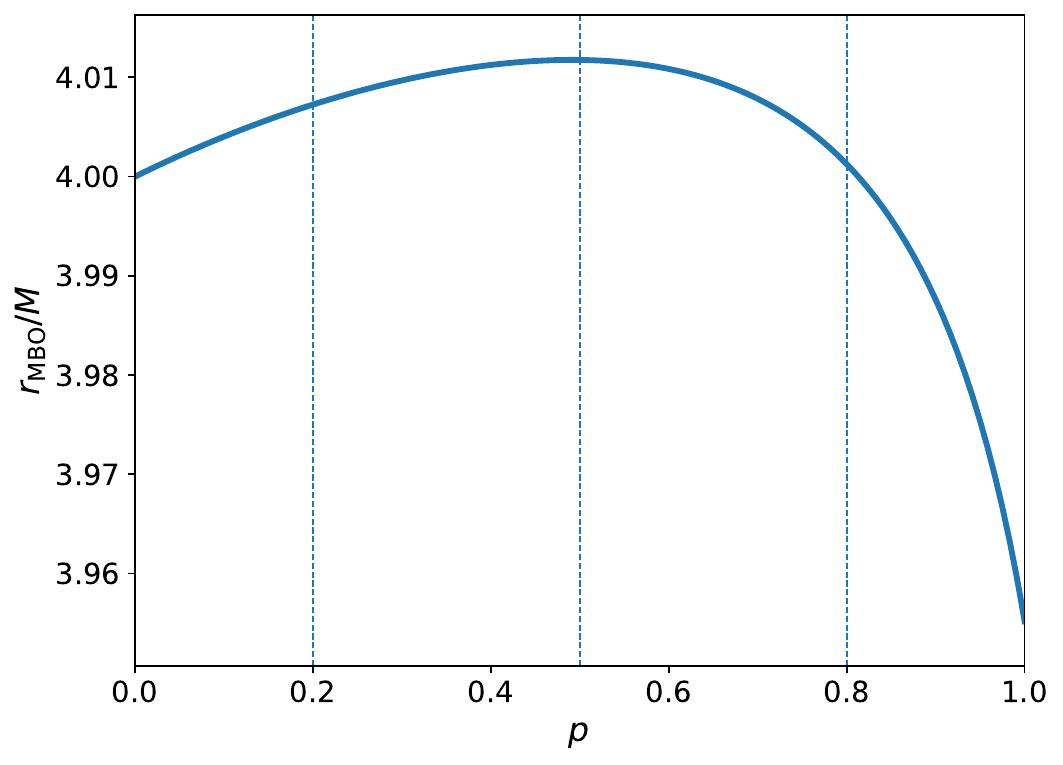}
    \includegraphics[width=0.45\linewidth]{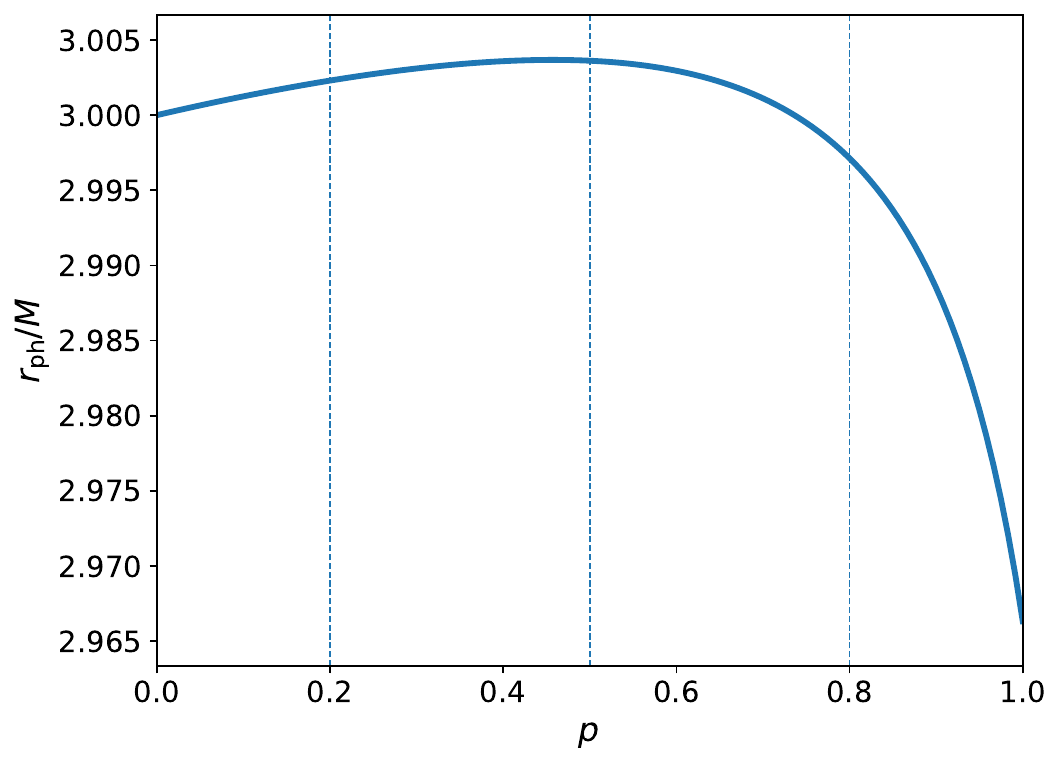}
    \caption{Characteristic radii of the static EsGB black hole as functions of the deformation parameter $p$.}
    \label{fig:h_i_m_p}
\end{figure*}

Figure~\ref{fig:h_i_m_p} shows how the main characteristic length scales of the EsGB black hole change when the deformation parameter $p$ is varied. These radii have different physical meanings: the horizon radius defines the boundary of the black hole, the ISCO marks the innermost radius of stable circular motion, the marginally bound orbit separates bound and unbound timelike circular trajectories, and the photon sphere determines the circular motion of massless particles. Together, they describe the structure of the strong-gravity region around the black hole.

The figure indicates that all characteristic radii recover their Schwarzschild values at $p=0$, namely $r_h/M=2$, $r_{\rm ISCO}/M=6$, $r_{\rm MBO}/M=4$, and $r_{\rm ph}/M=3$. As $p$ increases, these radii are shifted only moderately, showing that the EsGB correction does not destroy the usual hierarchy of black-hole length scales. Instead, it introduces a controlled deformation of the near-horizon geometry.

This behaviour has a clear physical interpretation. Since the scalar--Gauss--Bonnet term becomes more relevant in regions of high curvature, its main effect is expected close to the black hole. The shifts in $r_{\rm ISCO}$ and $r_{\rm MBO}$ show that the allowed region for stable and bound massive-particle motion is slightly modified. This is important for accretion physics, because the ISCO is commonly associated with the inner edge of a thin accretion disk. Similarly, the change in the photon-sphere radius indicates that light propagation and the optical appearance of the black hole can also be affected by the parameter $p$.

Therefore, Fig.~\ref{fig:h_i_m_p} provides a compact summary of how the EsGB deformation modifies the characteristic orbital structure of the spacetime. The corrections remain small but physically meaningful: they appear mainly in the strong-field region, where accretion dynamics, epicyclic frequencies, QPO generation, and photon trajectories are most sensitive to deviations from the Schwarzschild geometry.

\section{Epicyclic frequencies and geodesic HF-QPO models}
\label{sec:qpo_section}

Having established the structure of circular timelike motion in the static EsGB spacetime, we now turn to the corresponding orbital and epicyclic frequencies. These quantities provide the basis for the standard geodesic interpretation of HF-QPOs, since they connect the strong-field geometry directly to observable timing features. In the present framework, all frequencies depend on the black-hole mass $M$ and on the effective EsGB deformation parameter $p$, because the underlying metric functions $N(r)$ and $B(r)$ are fixed once the pair $(M,p)$ is specified \cite{Konoplya2020}.

\subsection{Orbital and epicyclic frequencies}

For a circular timelike geodesic at radius $r=r_c$, the orbital angular frequency measured at infinity is
\begin{equation}
\Omega_\phi(r_c)
=
\frac{d\phi}{dt}
=
\sqrt{\frac{N'(r_c)}{2r_c}}.
\label{eq:orbital_frequency_p}
\end{equation}
Since the spacetime is static and spherically symmetric, the vertical epicyclic frequency coincides with the orbital one,
\begin{equation}
\Omega_\theta(r_c)=\Omega_\phi(r_c).
\label{eq:vertical_frequency_p}
\end{equation}
Thus, in the static EsGB case, the nontrivial timing information is carried by the radial epicyclic sector.

The radial epicyclic frequency is obtained by perturbing a circular orbit according to
\begin{equation}
r(t)=r_c+\delta r(t),
\end{equation}
and expanding the radial equation of motion to linear order. In terms of the effective potential introduced in Section~\ref{sec:circular_motion}, one may write
\begin{equation}
\Omega_r^2(r_c)
=
\frac{1}{2B(r_c)}
\left(\frac{N(r_c)}{E_c(r_c)}\right)^2
\left.
\frac{d^2V_{\rm eff}}{dr^2}
\right|_{r=r_c}.
\label{eq:radial_frequency_p}
\end{equation}
By construction, the ISCO is determined by the condition
\begin{equation}
\Omega_r^2(r_{\rm ISCO})=0.
\label{eq:isco_frequency_condition_p}
\end{equation}
Therefore, the radial epicyclic frequency is especially sensitive to the near-horizon deformation controlled by the parameter $p$.

To compare with observations, one introduces the physical frequencies in Hz:
\begin{equation}
\nu_i(r)
=
\frac{1}{2\pi}\frac{c^3}{GM_{\rm BH}}\Omega_i(r),
\qquad
i\in\{\phi,r,\theta\},
\label{eq:frequencies_hz_p}
\end{equation}
where $M_{\rm BH}$ is the physical black-hole mass. In the Schwarzschild limit, one recovers
\begin{equation}
\Omega_\phi^2=\Omega_\theta^2=\frac{M}{r^3},
\qquad
\Omega_r^2=\frac{M(r-6M)}{r^4},
\label{eq:schwarzschild_frequencies_p}
\end{equation}
which shows explicitly that the radial epicyclic frequency vanishes at $r=6M$.

\subsection{Relativistic precession model}

A widely used geodesic interpretation of HF-QPOs is
the relativistic precession (RP) model, in which the twin
frequencies are identified as~\cite{StellaVietri1998,StellaVietri1999}
\begin{equation}
\nu_{\rm U}=\nu_\phi,
\qquad
\nu_{\rm L}=\nu_\phi-\nu_r.
\label{eq:RP_model_p}
\end{equation}
In this picture, the upper HF-QPO is associated with the orbital motion, while the lower frequency is identified with the periastron-precession combination.

In the present static EsGB setting, the deformation parameter $p$ affects both $\nu_\phi$ and $\nu_r$, but the influence on $\nu_r$ is typically stronger because the radial epicyclic mode depends more sensitively on the local curvature of the effective potential. As a result, the RP track
\begin{equation}
\nu_{\rm U}=\nu_{\rm U}(\nu_{\rm L};p,M)
\end{equation}
is shifted away from the Schwarzschild prediction once $p\neq0$. This makes the RP model a natural diagnostic of the scalar--Gauss--Bonnet deformation in the timing sector.

\subsection{Epicyclic resonance model}

A second important geodesic framework is the epicyclic
resonance (ER) model, where one identifies~\cite{AbramowiczKluzniak2001,KluzniakAbramowicz2001}
\begin{equation}
\nu_{\rm U}=\nu_\theta,
\qquad
\nu_{\rm L}=\nu_r,
\label{eq:ER_model_p}
\end{equation}
together with the resonance condition
\begin{equation}
\frac{\nu_\theta}{\nu_r}=\frac{3}{2}.
\label{eq:resonance_condition_p}
\end{equation}
Because the present spacetime is static and spherically symmetric, one has $\nu_\theta=\nu_\phi$, and therefore the resonance condition reduces to
\begin{equation}
\frac{\nu_\phi}{\nu_r}=\frac{3}{2}.
\label{eq:resonance_static_p}
\end{equation}
The corresponding resonance radius is therefore a $p$-dependent quantity,
\begin{equation}
r_{3:2}=r_{3:2}(p),
\end{equation}
and the associated HF-QPO pair also changes systematically with the EsGB deformation.

\begin{table}[t]
\caption{Resonance radius for the $3{:}2$ frequency ratio.}
\label{tab:resonance_radius_p}
\centering
\renewcommand{\arraystretch}{1.12}
\begin{tabular}{cc}
\toprule
$p$ & $r_{3:2}/M$ \\
\midrule
0.0 & 10.800 \\
0.2 & 10.815 \\
0.5 & 10.823 \\
0.8 & 10.805 \\
\bottomrule
\end{tabular}
\end{table}

Table~\ref{tab:resonance_radius_p} shows the radius satisfying the resonance condition $\nu_\phi/\nu_r=3/2$ for selected values of the EsGB deformation parameter. The Schwarzschild value is recovered at $p=0$, while nonzero values of $p$ produce only small but nonmonotonic shifts in $r_{3:2}$. This indicates that the scalar--Gauss--Bonnet correction changes the orbital and radial epicyclic frequencies in a non-identical way. Since the $3{:}2$ ratio plays an important role in many interpretations of high-frequency QPOs, the displacement of $r_{3:2}$ provides a useful measure of how the EsGB deformation can affect the region where resonant oscillations are generated.

It should be emphasized that the resonance radius $r_{3:2}$ listed in Table~\ref{tab:resonance_radius_p} is not the same quantity as the fitted emission radius used in the RP-model MCMC analysis. The former is obtained from the independent condition $\nu_\phi/\nu_r=3/2$, whereas the latter is determined by the observed pair through $\nu_U=\nu_\phi$ and $\nu_L=\nu_\phi-\nu_r$. Therefore, the difference between $r_{3:2}/M\simeq 10.8$ and the RP fitted radii around $r/M\simeq 6.7$ is expected.

\subsection{Phenomenological role of the deformation parameter}

The formulas above show that the static EsGB timing sector is controlled by the same deformation parameter that determines the background geometry. Once the branch is fixed, increasing $p$ modifies the metric functions, which in turn shifts the effective potential, the characteristic circular-orbit radii, and finally the orbital and epicyclic frequencies. In practical terms, the strongest imprint of the EsGB sector is expected to appear in the radial epicyclic frequency and in the quantities derived from it, namely the lower RP frequency and the ER resonance condition.

This observation is particularly useful for phenomenology. Rather than attempting to constrain the continued-fraction coefficients $\epsilon$, $a_i$, and $b_i$ directly, one may work with the smaller and more physical parameter set
\begin{equation}
(M,p,r),
\end{equation}
where $M$ is the black-hole mass, $p$ is the effective EsGB deformation parameter, and $r$ is the relevant orbital or resonance radius. In this way, the static EsGB timing problem becomes directly comparable to observational HF-QPO analyses.

\subsection{Remarks on the strong-coupling regime}

Finally, one should keep in mind that the regime $p\to1$ corresponds to the maximal-coupling limit of the static EsGB family. In this region the continued-fraction approximation converges more slowly, and large couplings are likely to be associated with unstable configurations \cite{Konoplya2020}. For this reason, the most reliable phenomenological analysis is expected to arise from the nonextremal regime, where the analytic parametrization remains accurate and the background geometry is physically better motivated.

The next step is therefore to confront the theoretical frequency relations derived above with observed HF-QPO data and examine how the pair $(M,p)$ can be constrained from timing measurements.

\begin{figure*}[ht!]
    \centering
    \includegraphics[width=0.45\linewidth]{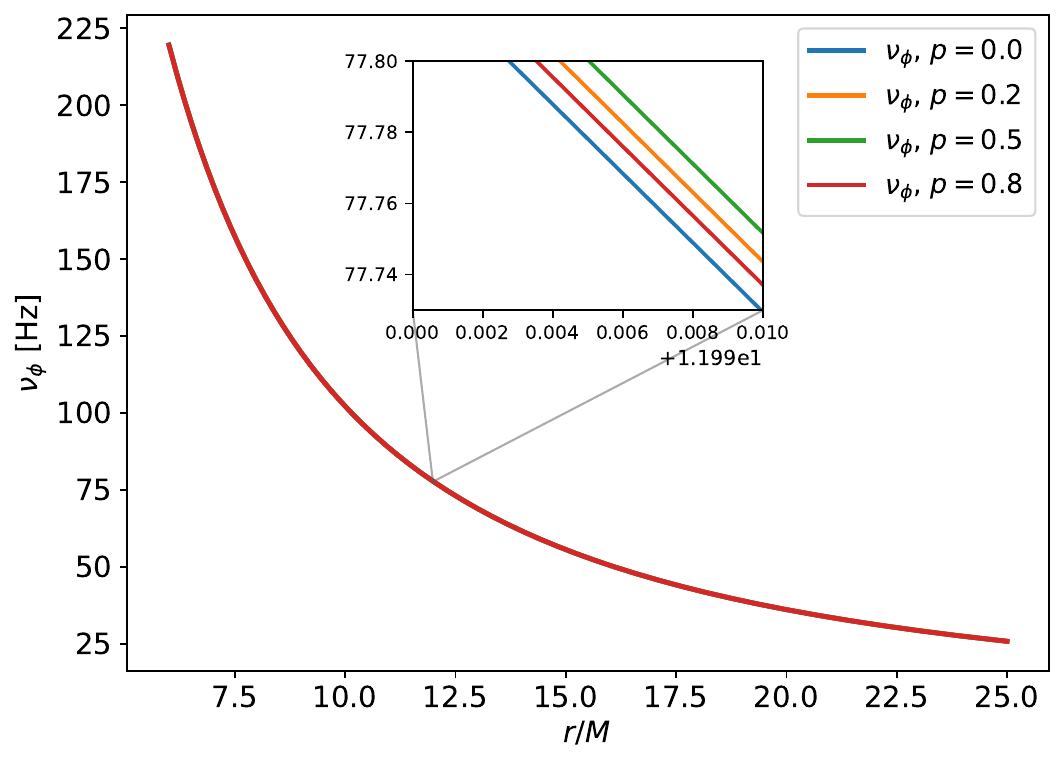}
    \includegraphics[width=0.45\linewidth]{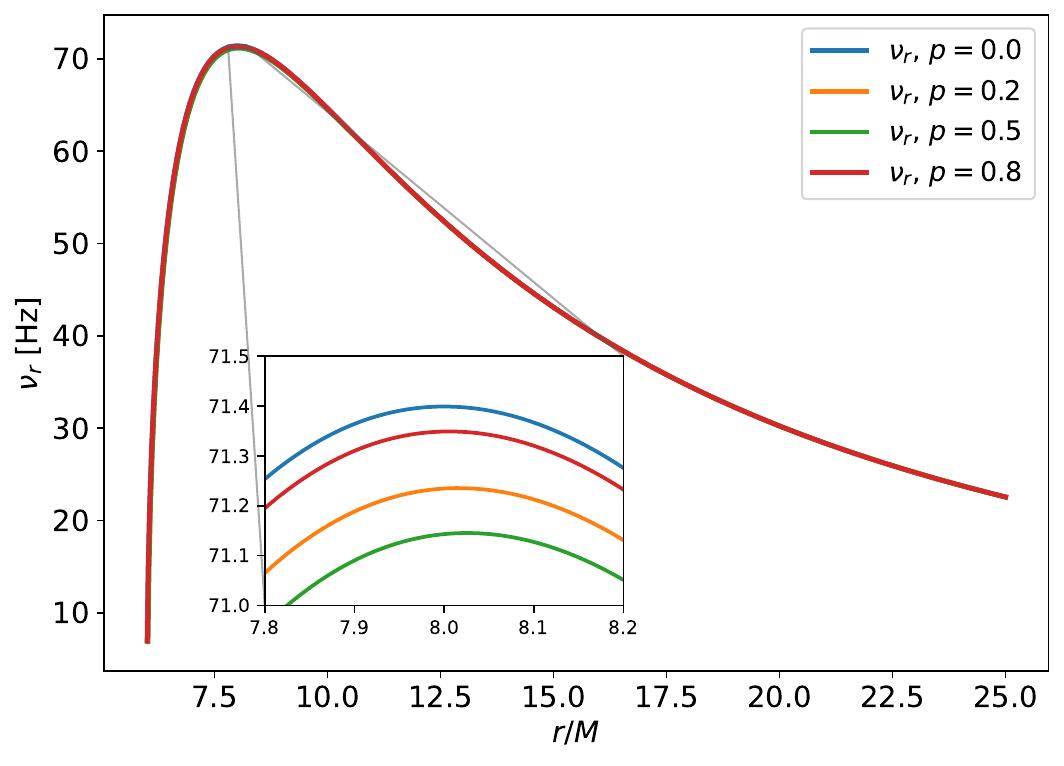}
    \includegraphics[width=0.45\linewidth]{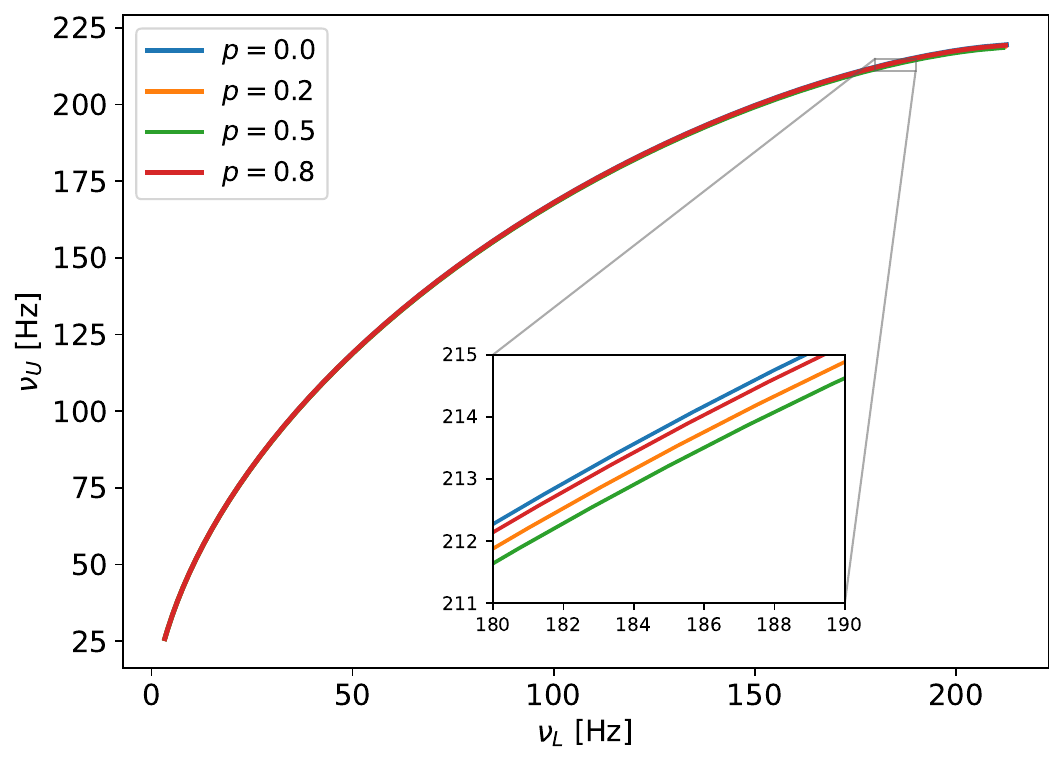}
    \includegraphics[width=0.45\linewidth]{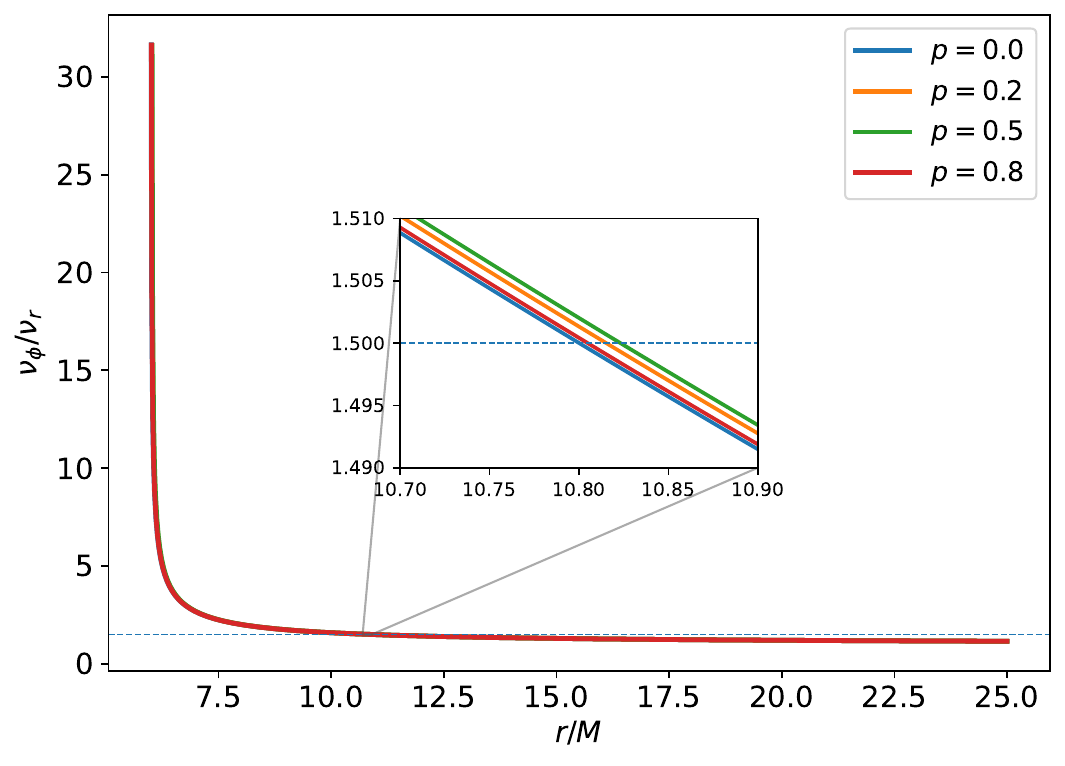}
    \caption{Orbital and epicyclic frequency profiles, RP-model tracks, and the $3{:}2$ frequency ratio for $M=10M_\odot$.}
    \label{fig:frequency}
\end{figure*}

Figure~\ref{fig:frequency} shows the behaviour of the orbital frequency $\nu_\phi$, the radial epicyclic frequency $\nu_r$, the corresponding relativistic-precession-model frequency tracks, and the ratio $\nu_\phi/\nu_r$ for a representative black hole mass $M=10M_\odot$. These frequencies are directly connected with the motion of matter in the inner accretion region. In particular, $\nu_\phi$ describes the azimuthal orbital motion, while $\nu_r$ measures small radial oscillations around a circular orbit.

The upper panels show that the orbital frequency decreases with increasing radius, as expected for particles moving farther from the black hole. The effect of the EsGB parameter $p$ on $\nu_\phi$ is relatively mild. This indicates that the azimuthal orbital motion is not strongly modified by the deformation, at least in the radial range considered here. By contrast, the radial epicyclic frequency is more sensitive to the change in $p$. Since $\nu_r$ is controlled by the curvature of the effective potential near circular orbits, even a small deformation of the near-horizon geometry can noticeably affect radial oscillations.

The RP-model panel translates these frequency profiles into observable QPO pairs. In this model, the upper frequency is associated with the orbital motion, while the lower frequency is related to the difference between the orbital and radial epicyclic frequencies. Therefore, any change in $\nu_r$ directly shifts the predicted QPO tracks. This makes the radial epicyclic frequency an important diagnostic quantity for testing deviations from the Schwarzschild geometry.

The ratio $\nu_\phi/\nu_r$ is especially relevant because high-frequency QPOs are often observed close to simple rational ratios, particularly the $3{:}2$ ratio. The lower-right panel shows how the radius satisfying this ratio changes with the EsGB parameter. A shift of this radius means that the region where resonant or quasi-resonant oscillations can occur is displaced by the scalar--Gauss--Bonnet correction. Thus, Fig.~\ref{fig:frequency} demonstrates that timing observables, especially those involving $\nu_r$, can provide a sensitive probe of the strong-field deformation parameter $p$.

\section{Observational constraints from QPO data}
\label{sec:mcmc_constraints}
Although we refer to the results as observational constraints, they should be understood as model-dependent phenomenological constraints within the static EsGB RP framework. In particular, the inferred ranges of $(M,p,r)$ depend on the adopted geodesic frequency identification, the static approximation, and the chosen prior structure. Therefore, the results should not be interpreted as definitive measurements of the EsGB coupling, but rather as compatibility regions selected by the observed twin-peak QPO pairs.

The previous sections show how the static EsGB geometry determines the geodesic frequencies. We now use this connection to compare the model with observed twin-peak HF-QPOs. In this first source-by-source analysis we adopt the RP identification,
\begin{equation}
\nu_{\rm U}^{\rm th}=\nu_\phi(M,p,r),
\qquad
\nu_{\rm L}^{\rm th}=\nu_\phi(M,p,r)-\nu_r(M,p,r),
\label{eq:rp_theory_p_revised}
\end{equation}
and we fit the reduced parameter vector
\begin{equation}
\bm{\Theta}=(M,p,r).
\label{eq:parameter_vector_p_revised}
\end{equation}
Here $M$ is measured in solar masses, $p$ is the EsGB deformation parameter, and $r$ is the orbital radius in units of $M$.

\subsection{Observed QPO pairs}

The observational data used in the fit are listed in Table~\ref{tab:qpo_data}. The stellar-mass black-hole binaries XTE J1550$-$564, GRO J1655$-$40, and GRS 1915+105 are standard HF-QPO sources, while M82 X-1 is included as an intermediate-mass black-hole candidate with a stable twin-peak QPO pair. The values are used as phenomenological inputs for testing the static EsGB timing model.

\begin{table*}[t]
\caption{Observed twin-peak QPO pairs used in the RP-model MCMC analysis.}
\label{tab:qpo_data}
\centering
\renewcommand{\arraystretch}{1.12}
\begin{tabular}{lccccc}
\toprule
Source & $\nu^{\rm obs}_{U}$ [Hz] & $\sigma_U$ [Hz] 
& $\nu^{\rm obs}_{L}$ [Hz] & $\sigma_L$ [Hz] & Reference \\
\midrule
XTE J1550$-$564 & 276 & 3 & 184 & 5 & \cite{Remillard2002} \\
GRO J1655$-$40 & 450 & 3 & 300 & 5 & \cite{Strohmayer2001} \\
GRS 1915+105 & 168 & 3 & 113 & 5 & \cite{RemillardMcClintock2006} \\
M82 X-1 & 5.07 & 0.06 & 3.32 & 0.06 & \cite{Pasham2014} \\
\bottomrule
\end{tabular}
\end{table*}

\subsection{Likelihood and priors}

For each source, we assume independent Gaussian uncertainties for the two observed QPO frequencies. The chi-square function is
\begin{equation}
\begin{aligned}
\chi^2(M,p,r)
={}&
\frac{
\left[\nu_{\rm U}^{\rm th}(M,p,r)
-\nu_{\rm U}^{\rm obs}\right]^2
}{\sigma_{\rm U}^2}
\\
&+
\frac{
\left[\nu_{\rm L}^{\rm th}(M,p,r)
-\nu_{\rm L}^{\rm obs}\right]^2
}{\sigma_{\rm L}^2},
\end{aligned}
\label{eq:chi2_single_p}
\end{equation}
with likelihood
\begin{equation}
\mathcal{L}(M,p,r)\propto
\exp\left[-\frac{\chi^2(M,p,r)}{2}\right].
\label{eq:likelihood_p}
\end{equation}
The posterior distribution is therefore
\begin{equation}
P(M,p,r\mid {\rm data})
\propto
\mathcal{L}(M,p,r)\,\Pi_M(M)\,\Pi_p(p)\,\Pi_r(r),
\label{eq:posterior_p}
\end{equation}
where the prior factors are restricted to the source-dependent top-hat intervals listed in Table~\ref{tab:source_priors_p}.

Each source supplies two measured QPO frequencies, whereas the fitted vector contains three correlated parameters, $(M,p,r)$. The QPO-only problem is therefore intrinsically underconstrained. In the revised primary analysis, we adopt the physically allowed uniform prior
\begin{equation}
\Pi_p(p)=\mathcal{U}(0,1),
\label{eq:uniform_p_prior}
\end{equation}
while retaining the same Gaussian localization factors for $M$ and $r/M$ as in the original analysis. These factors restrict the chains to the frequency-matching region and should be interpreted as phenomenological regularization terms rather than independent astrophysical measurements.

 The central values and widths of the original Gaussian priors on $p$ were obtained as the sample means and standard deviations of the viable parameter points retained in a preliminary numerical frequency-matching scan over the source-dependent top-hat domains. They were introduced solely as phenomenological localization values for the partially degenerate fit and were not derived from independent physical or observational information.

To assess the influence of the prior assigned to the EsGB deformation parameter, we repeat the complete source-by-source analysis using the original truncated Gaussian prior
\begin{equation}
\Pi_p^{\rm G}(p)\propto
\exp\left[-\frac{(p-\mu_p)^2}{2\sigma_p^2}\right],
\qquad 0\leq p\leq1.
\label{eq:gaussian_p_prior}
\end{equation}
The QPO likelihood, observational uncertainties, top-hat supports, Gaussian localization factors for $M$ and $r/M$, physical-orbit conditions, and sampler settings are identical in the two runs. Consequently, the comparison isolates the sensitivity of the inference to the prior on $p$.

\begin{table*}[t]
\caption{Source-dependent parameter supports and priors used in the revised static EsGB RP analysis. The uniform prior on $p$ defines the primary analysis, whereas the original truncated Gaussian prior on $p$ is used only for the controlled sensitivity test.}
\label{tab:source_priors_p}
\centering
\scriptsize
\renewcommand{\arraystretch}{1.15}
\resizebox{\textwidth}{!}{%
\begin{tabular}{lccccc}
\toprule
Source
& Top-hat $M/M_\odot$
& Top-hat $r/M$
& Gaussian localization for $M/M_\odot$
& Gaussian localization for $r/M$
& Prior on $p$: primary; sensitivity \\
\midrule
XTE J1550$-$564
& $[6.48,6.88]$
& $[6.62,6.88]$
& $6.669\pm0.090$
& $6.754\pm0.053$
& $\mathcal{U}(0,1)$; $\mathcal{N}_{[0,1]}(0.495,0.296^2)$ \\
GRO J1655$-$40
& $[4.02,4.16]$
& $[6.67,6.84]$
& $4.096\pm0.035$
& $6.756\pm0.037$
& $\mathcal{U}(0,1)$; $\mathcal{N}_{[0,1]}(0.507,0.274^2)$ \\
GRS 1915+105
& $[10.4,11.6]$
& $[6.50,7.04]$
& $11.009\pm0.258$
& $6.734\pm0.083$
& $\mathcal{U}(0,1)$; $\mathcal{N}_{[0,1]}(0.486,0.281^2)$ \\
M82 X-1
& $[338,378]$
& $[6.70,6.97]$
& $358.126\pm9.270$
& $6.827\pm0.080$
& $\mathcal{U}(0,1)$; $\mathcal{N}_{[0,1]}(0.480,0.274^2)$ \\
\bottomrule
\end{tabular}}
\end{table*}

\subsection{MCMC setup and convergence}

Posterior sampling was performed with the affine-invariant ensemble sampler implemented in \texttt{emcee}~\cite{ForemanMackey2013}. For each source and each choice of the prior on $p$, we used 16 walkers and evolved every walker for $10^{5}$ steps. The first $2\times10^{4}$ steps were discarded as burn-in, and every 25th post-burn-in state was retained. This procedure yields 51,200 retained post-burn-in samples for each source and each prior choice.

For the primary uniform-prior runs, the mean acceptance fractions lie between 0.596 and 0.599. The integrated autocorrelation times range from approximately 46.5 to 59.0 steps, and the post-burn-in chains are longer than $1.3\times10^{3}$ autocorrelation times for every parameter. The corresponding Gaussian-prior sensitivity runs have mean acceptance fractions between 0.636 and 0.639 and integrated autocorrelation times of approximately 41.7--45.8 steps. The trace plots were also inspected and showed no persistent drift after burn-in. These diagnostics indicate stable sampling and demonstrate that the broad posterior of $p$ is not a consequence of insufficient chain length or poor convergence.

\subsection{Posterior estimates with the uniform prior}

Table~\ref{tab:bestfit_p} presents the posterior medians and central $68\%$ credible intervals obtained using $p\sim\mathcal{U}(0,1)$. The fitted values of $M$ and $r/M$ remain localized, whereas the marginal posterior of $p$ spans most of the physically allowed interval for every source.

\begin{table*}[t]
\caption{Posterior estimates for the primary static EsGB RP analysis with $p\sim\mathcal{U}(0,1)$. The quoted values are posterior medians with central $68\%$ credible intervals. They represent model-dependent compatibility regions rather than independent measurements of the EsGB deformation.}
\label{tab:bestfit_p}
\centering
\renewcommand{\arraystretch}{1.14}
\begin{tabular*}{\textwidth}{@{\extracolsep{\fill}} lcccc}
\toprule
Source & $M/M_\odot$ & $p$ & $r/M$ & Mean acceptance \\
\midrule
XTE J1550$-$564
& $6.6699^{+0.0693}_{-0.0672}$
& $0.4916^{+0.3432}_{-0.3354}$
& $6.7562^{+0.0378}_{-0.0375}$
& 0.5977 \\
GRO J1655$-$40
& $4.0939^{+0.0264}_{-0.0266}$
& $0.4922^{+0.3362}_{-0.3344}$
& $6.7548^{+0.0249}_{-0.0247}$
& 0.5965 \\
GRS 1915+105
& $11.0086^{+0.1899}_{-0.1877}$
& $0.4999^{+0.3358}_{-0.3424}$
& $6.7345^{+0.0609}_{-0.0595}$
& 0.5993 \\
M82 X-1
& $357.7526^{+5.5544}_{-5.4449}$
& $0.4982^{+0.3326}_{-0.3370}$
& $6.8231^{+0.0472}_{-0.0472}$
& 0.5971 \\
\bottomrule
\end{tabular*}
\end{table*}

\begin{figure*}[t]
\centering
\begin{minipage}{0.48\textwidth}
\centering
\includegraphics[width=\linewidth]{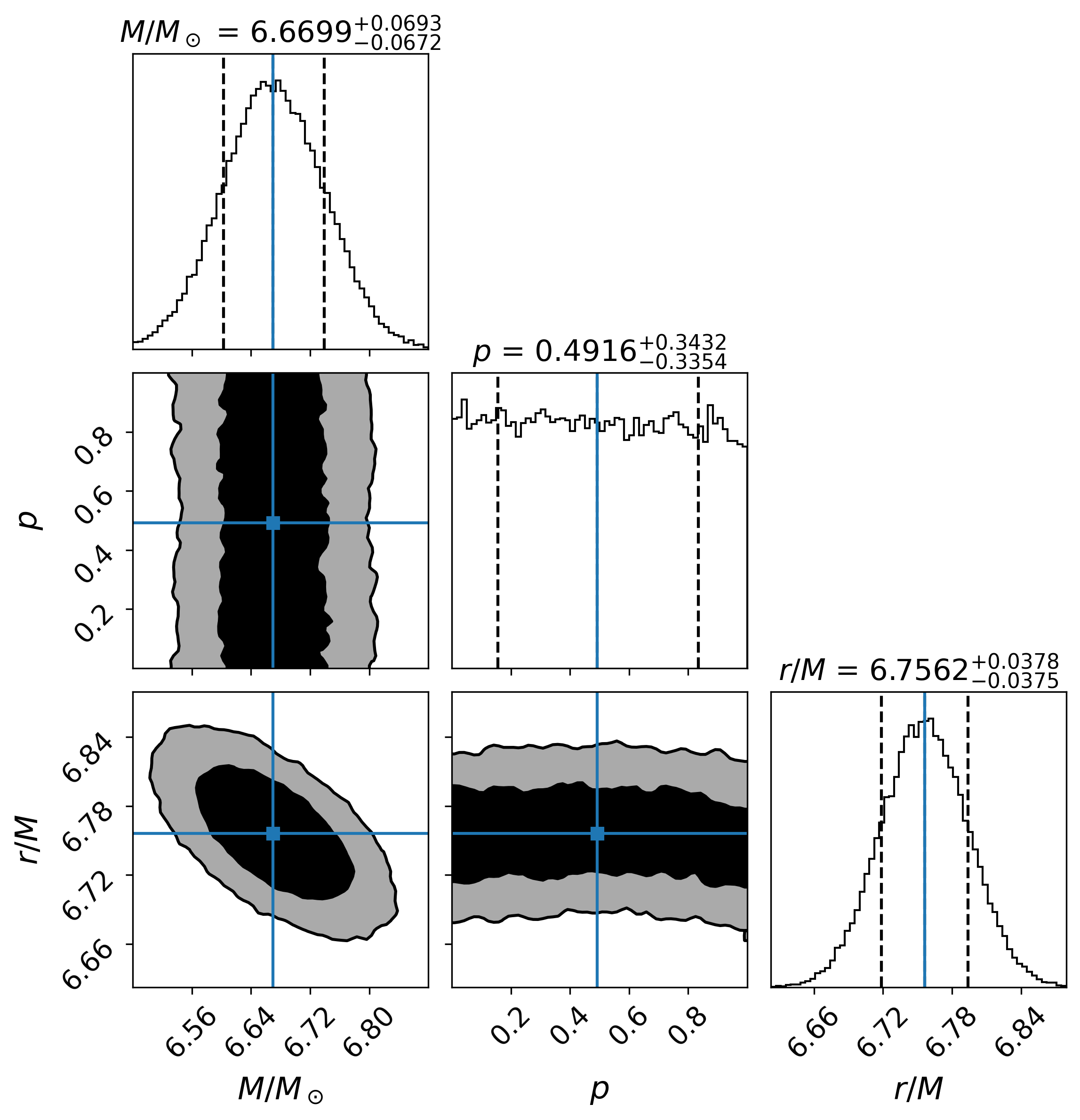}\\[-2mm]
\textbf{(a) XTE J1550$-$564}
\end{minipage}\hfill
\begin{minipage}{0.48\textwidth}
\centering
\includegraphics[width=\linewidth]{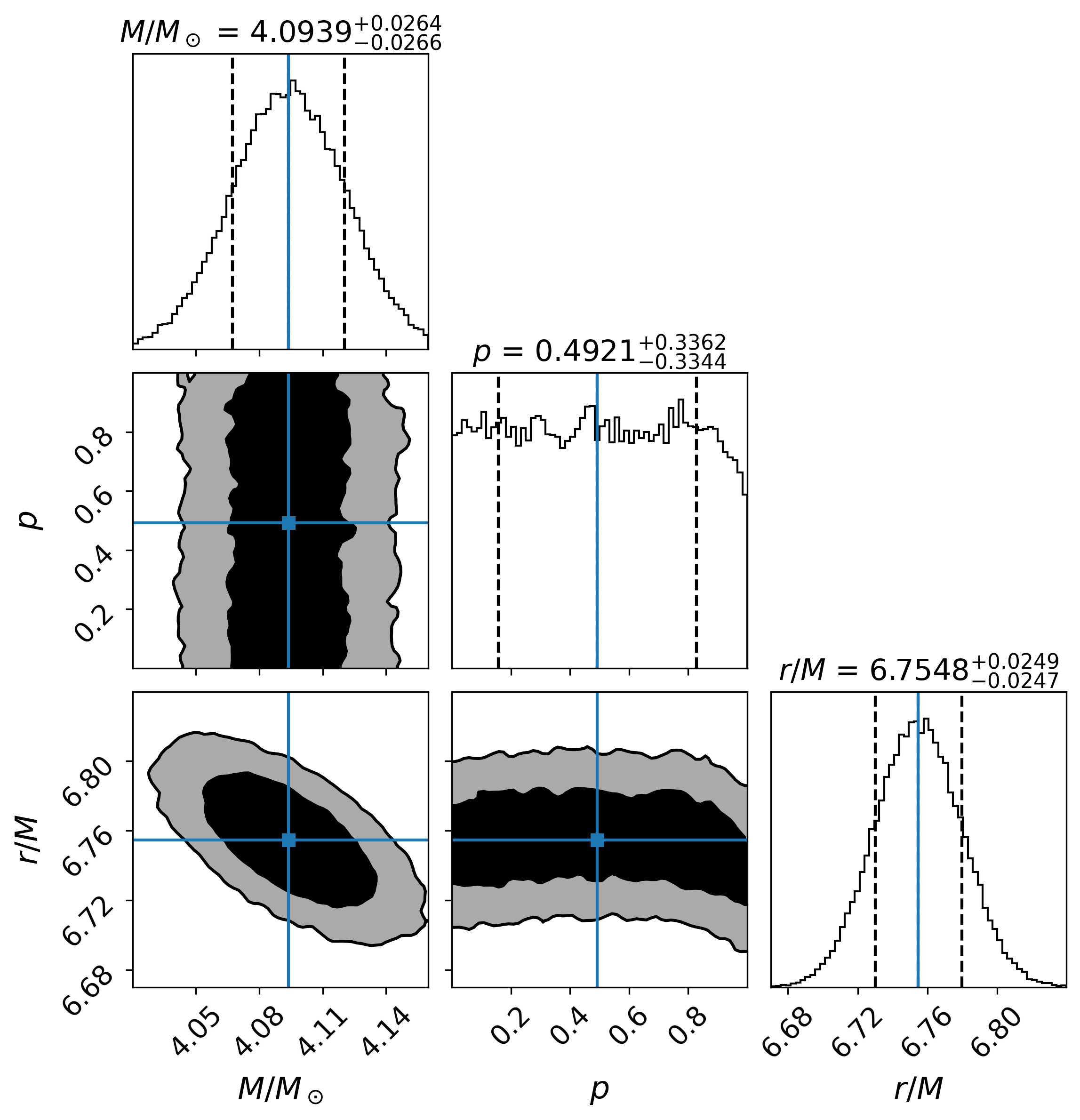}\\[-2mm]
\textbf{(b) GRO J1655$-$40}
\end{minipage}

\vspace{2mm}

\begin{minipage}{0.48\textwidth}
\centering
\includegraphics[width=\linewidth]{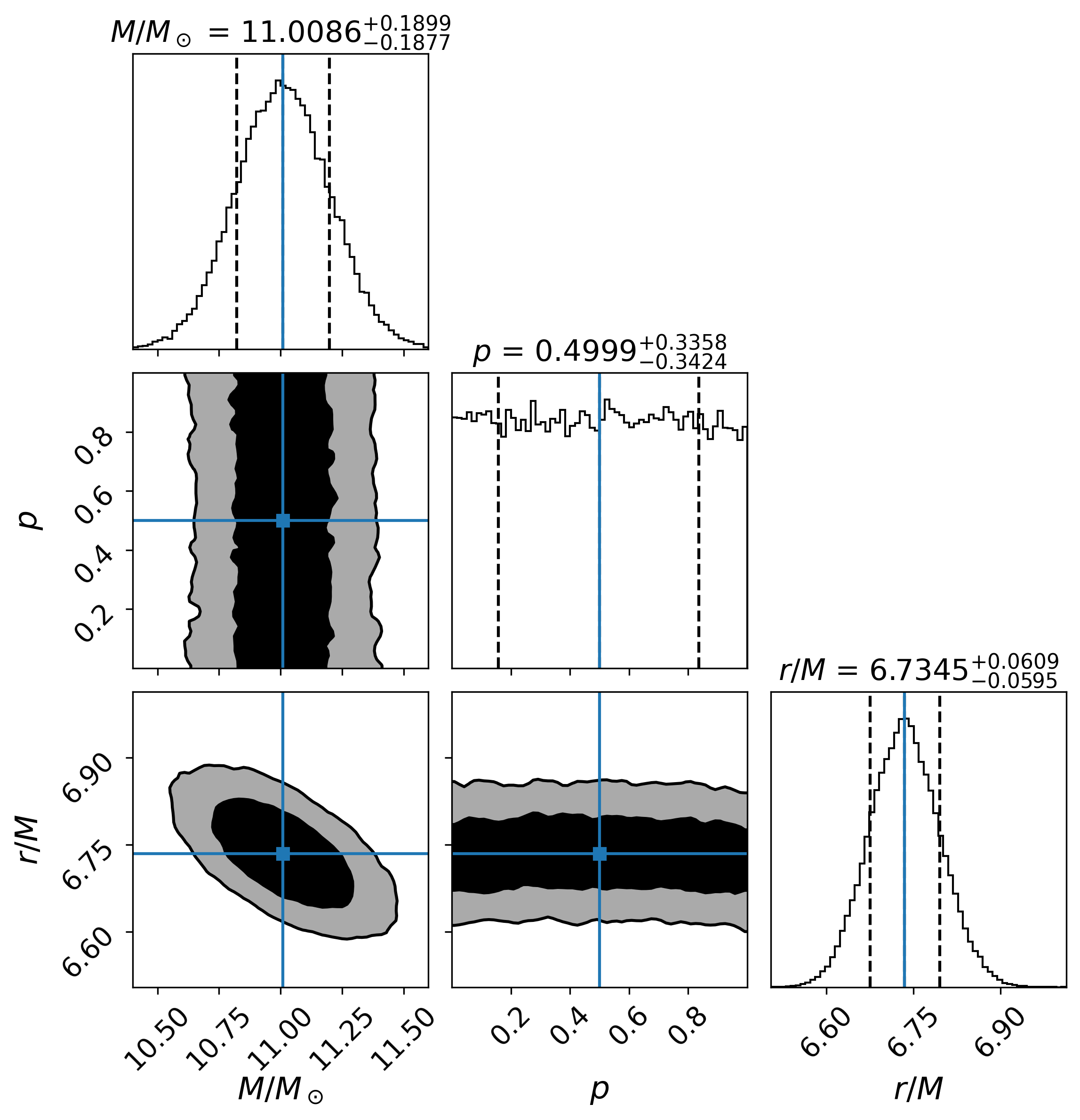}\\[-2mm]
\textbf{(c) GRS 1915+105}
\end{minipage}\hfill
\begin{minipage}{0.48\textwidth}
\centering
\includegraphics[width=\linewidth]{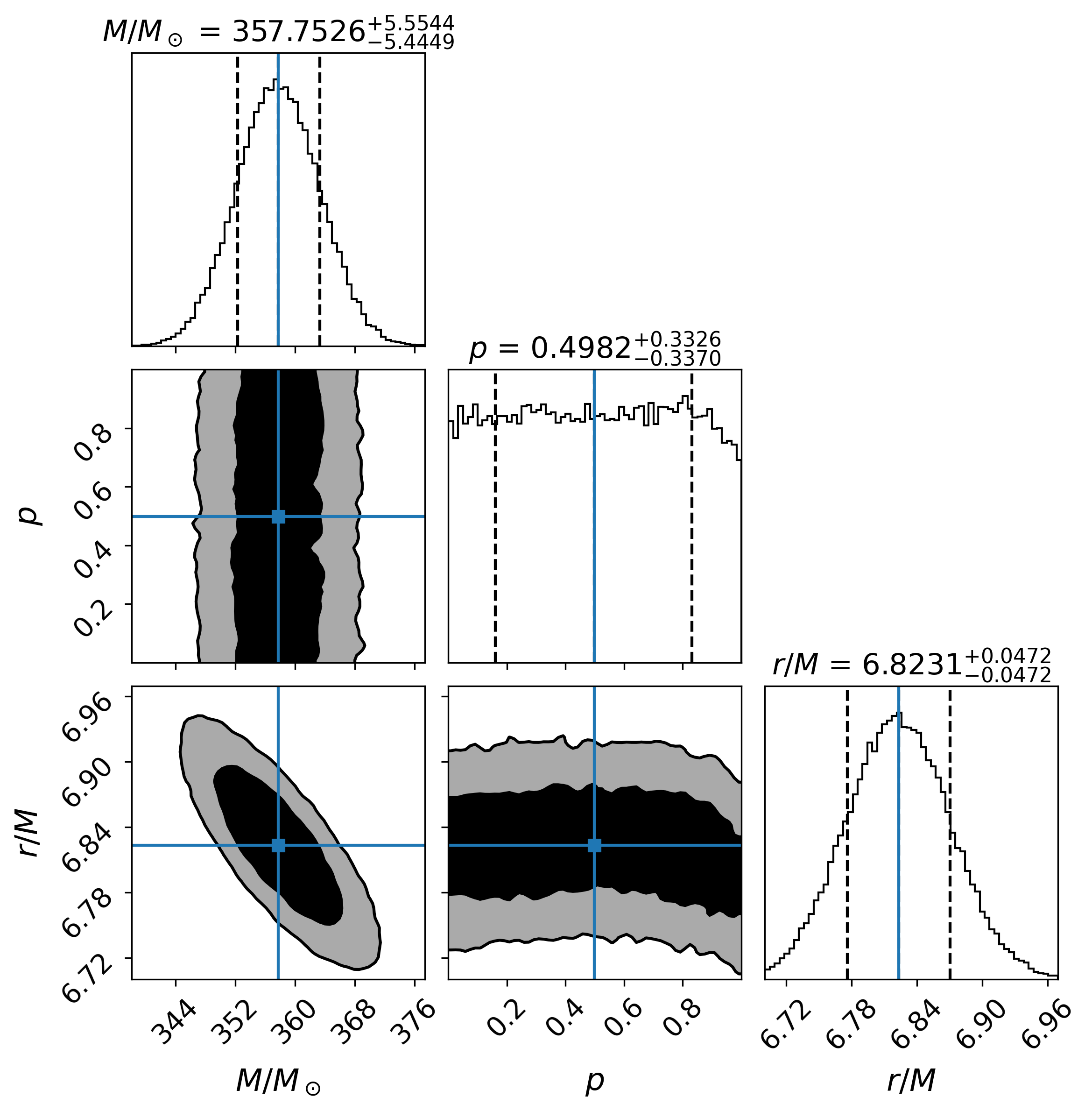}\\[-2mm]
\textbf{(d) M82 X-1}
\end{minipage}
\caption{Marginalized one- and two-dimensional posterior distributions of $(M,p,r/M)$ obtained using the uniform prior $p\sim\mathcal{U}(0,1)$. For every source, the marginal posterior of $p$ remains broad over nearly the complete physical interval, while $M$ and $r/M$ exhibit a clear anticorrelation. The observed QPO pair therefore constrains correlated combinations of $(M,p,r)$ but provides little independent information about the EsGB deformation parameter.}
\label{fig:MCMC}
\end{figure*}

The posterior values of $M$ should be understood as effective mass scales selected by the static RP-frequency prescription, rather than as revised dynamical masses of the corresponding sources. The uniform-prior corner plots in Fig.~\ref{fig:MCMC} show a clear anticorrelation between $M$ and $r/M$. In contrast, the $p$ direction remains broad, indicating that changes in the deformation can be compensated by small shifts of the mass scale and emission radius.

The theoretical QPO frequencies evaluated at the posterior medians are compared with the observations in Table~\ref{tab:qpo_fit_p}. All four observed pairs are reproduced within their adopted uncertainties, showing that the static EsGB RP prescription can accommodate the selected QPO pairs at the phenomenological level. This agreement, however, does not imply that $p$ is independently measured.

\begin{table*}[t]
\caption{Observed QPO frequencies and theoretical RP-model values evaluated at the posterior medians of the uniform-prior runs. The final column gives the QPO chi-square at the posterior median.}
\label{tab:qpo_fit_p}
\centering
\renewcommand{\arraystretch}{1.12}
\begin{tabular*}{\textwidth}{@{\extracolsep{\fill}} lccccc}
\toprule
Source
& $\nu_{\rm U}^{\rm obs}$ [Hz]
& $\nu_{\rm U}^{\rm th}$ [Hz]
& $\nu_{\rm L}^{\rm obs}$ [Hz]
& $\nu_{\rm L}^{\rm th}$ [Hz]
& $\chi^2_{\rm QPO}$ \\
\midrule
XTE J1550$-$564 & 276 & 276.072 & 184 & 184.886 & 0.032 \\
GRO J1655$-$40  & 450 & 449.924 & 300 & 301.439 & 0.083 \\
GRS 1915+105    & 168 & 168.074 & 113 & 113.294 & 0.004 \\
M82 X-1         & 5.07 & 5.071 & 3.32 & 3.330 & 0.030 \\
\bottomrule
\end{tabular*}
\end{table*}

\subsection{Prior sensitivity and identifiability of the deformation parameter}

\begin{figure*}[t]
\centering
\begin{minipage}{0.48\textwidth}
\centering
\includegraphics[width=\linewidth]{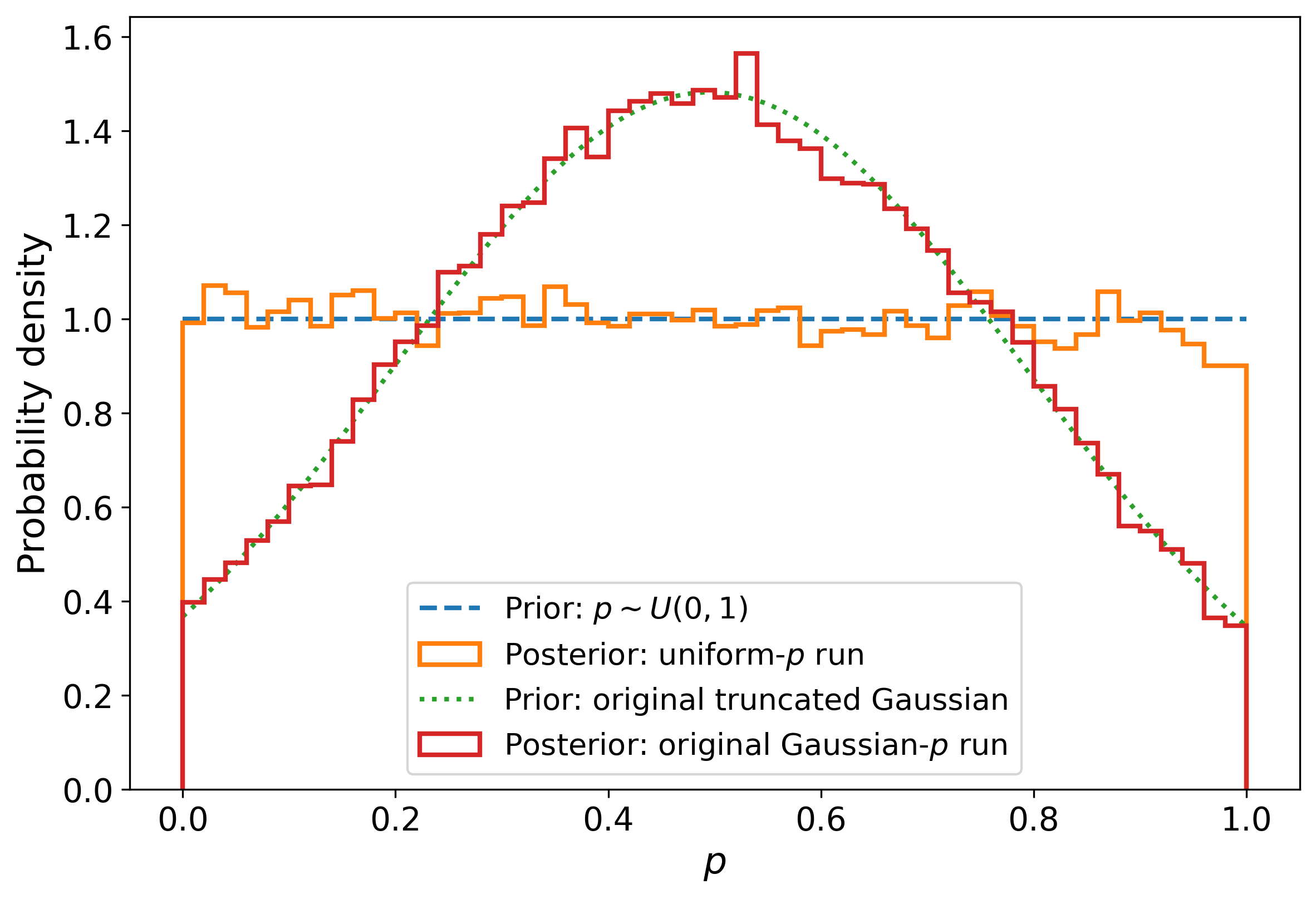}\\[-2mm]
\textbf{(a) XTE J1550$-$564}
\end{minipage}\hfill
\begin{minipage}{0.48\textwidth}
\centering
\includegraphics[width=\linewidth]{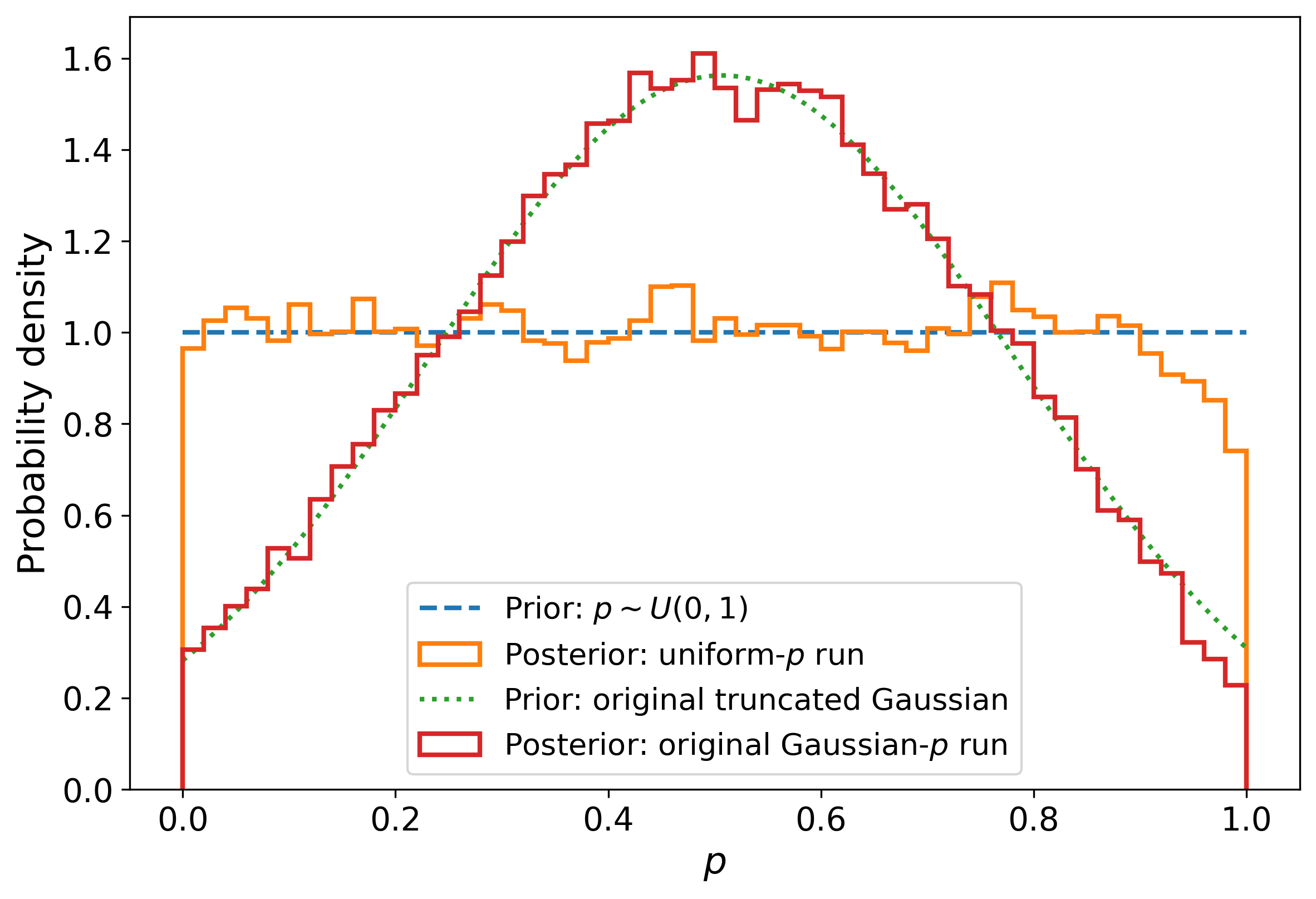}\\[-2mm]
\textbf{(b) GRO J1655$-$40}
\end{minipage}

\vspace{2mm}

\begin{minipage}{0.48\textwidth}
\centering
\includegraphics[width=\linewidth]{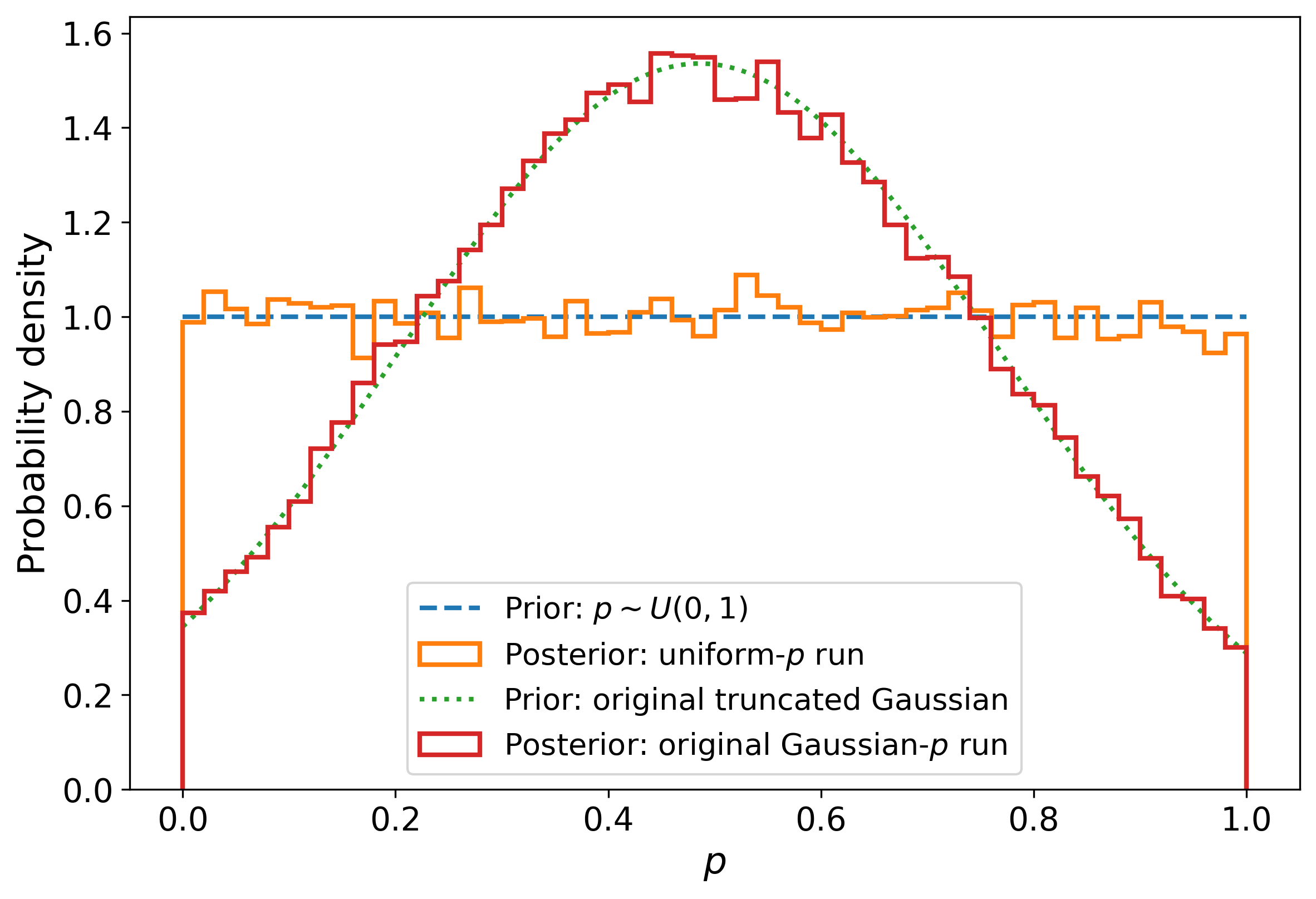}\\[-2mm]
\textbf{(c) GRS 1915+105}
\end{minipage}\hfill
\begin{minipage}{0.48\textwidth}
\centering
\includegraphics[width=\linewidth]{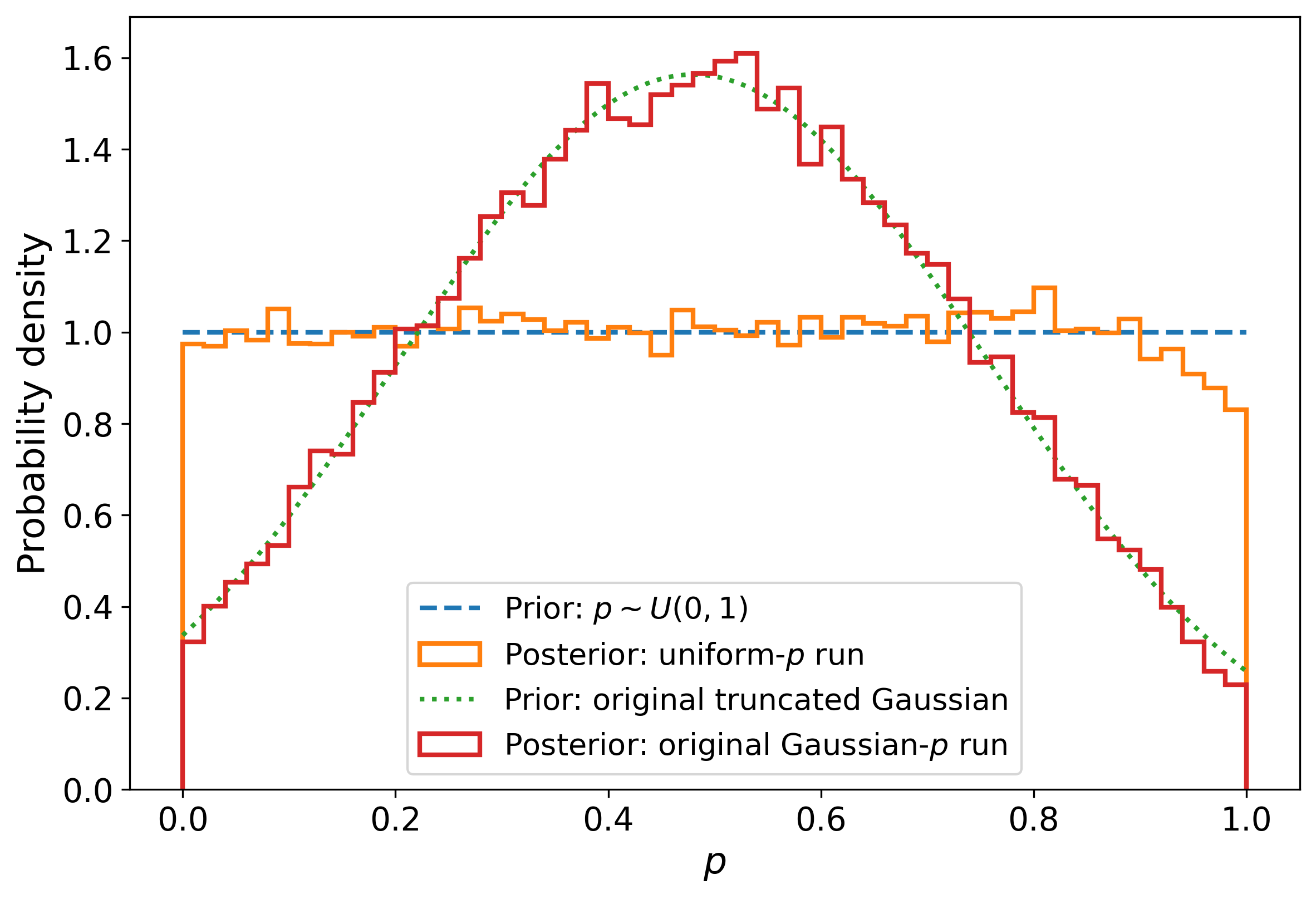}\\[-2mm]
\textbf{(d) M82 X-1}
\end{minipage}
\caption{Controlled prior-sensitivity analysis for the EsGB deformation parameter $p$. The uniform and original truncated-Gaussian priors are compared with the corresponding marginal posteriors. In all four sources, each posterior closely follows the adopted prior, demonstrating that the present QPO pairs provide negligible independent information about $p$.}
\label{fig:p_prior_sensitivity}
\end{figure*}

Figure~\ref{fig:p_prior_sensitivity} compares the prior and marginal posterior distributions of $p$ for the uniform and original truncated-Gaussian choices. In every source, the uniform-prior posterior remains almost flat, while the Gaussian-prior posterior closely follows the corresponding truncated Gaussian. Thus, the numerical location and width of the marginal posterior are controlled mainly by the adopted prior rather than by the QPO likelihood.

To quantify this result, we define the central $68\%$ width ratio
\begin{equation}
\mathcal{R}_{68}
=
\frac{p_{84}^{\rm post}-p_{16}^{\rm post}}
{p_{84}^{\rm prior}-p_{16}^{\rm prior}},
\label{eq:width_ratio_p}
\end{equation}
and the normalized median shift
\begin{equation}
\mathcal{S}_{p}
=
\frac{p_{50}^{\rm post}-p_{50}^{\rm prior}}
{\sigma_{p}^{\rm prior}}.
\label{eq:median_shift_p}
\end{equation}
The results are summarized in Table~\ref{tab:p_prior_sensitivity}. Values of $\mathcal{R}_{68}$ close to unity indicate negligible posterior contraction relative to the prior, while $|\mathcal{S}_{p}|\ll1$ indicates that the data do not displace the posterior median from the prior median.

\begin{table*}[t]
\caption{Controlled prior-sensitivity results for the EsGB deformation parameter. Here $\mathcal{R}_{68}$ is the posterior-to-prior central-width ratio defined in Eq.~\eqref{eq:width_ratio_p}, and $\mathcal{S}_{p}$ is the median shift in units of the prior standard deviation defined in Eq.~\eqref{eq:median_shift_p}.}
\label{tab:p_prior_sensitivity}
\centering
\small
\renewcommand{\arraystretch}{1.12}
\begin{tabular*}{\textwidth}{@{\extracolsep{\fill}} llccc}
\toprule
Source & Prior on $p$ & Posterior $p$ & $\mathcal{R}_{68}$ & $\mathcal{S}_{p}$ \\
\midrule
XTE J1550$-$564
& Uniform
& $0.4916^{+0.3432}_{-0.3354}$
& 0.9979 & $-0.0292$ \\
& Original Gaussian
& $0.4949^{+0.2621}_{-0.2571}$
& 1.0028 & $-0.0053$ \\
\addlinespace
GRO J1655$-$40
& Uniform
& $0.4922^{+0.3362}_{-0.3344}$
& 0.9862 & $-0.0272$ \\
& Original Gaussian
& $0.5022^{+0.2453}_{-0.2471}$
& 0.9946 & $-0.0150$ \\
\addlinespace
GRS 1915+105
& Uniform
& $0.4999^{+0.3358}_{-0.3424}$
& 0.9973 & $-0.0002$ \\
& Original Gaussian
& $0.4860^{+0.2553}_{-0.2494}$
& 1.0046 & $-0.0125$ \\
\addlinespace
M82 X-1
& Uniform
& $0.4982^{+0.3326}_{-0.3370}$
& 0.9848 & $-0.0062$ \\
& Original Gaussian
& $0.4860^{+0.2470}_{-0.2478}$
& 1.0002 & $+0.0094$ \\
\bottomrule
\end{tabular*}
\end{table*}

As a prior-independent diagnostic, we also calculated the QPO-only
profile chi-square. At each fixed value of $p$, it is defined as
\begin{equation}
\chi^2_{\rm prof}(p)
=
\min_{M,r}\chi^2(M,p,r),
\end{equation}
where the minimization is performed over the same source-dependent
top-hat supports and physical-orbit domain used in the MCMC analysis.
We then define
\begin{equation}
\Delta\chi^2_{\rm prof}(p)
=
\chi^2_{\rm prof}(p)
-
\min_{p}\chi^2_{\rm prof}(p).
\end{equation}

For the uniform-prior runs, $\mathcal{R}_{68}$ lies between 0.985 and 0.998, and the posterior medians differ from the uniform-prior median by less than 0.03 prior standard deviations. The original Gaussian-prior runs give the same qualitative result, with $\mathcal{R}_{68}$ between 0.995 and 1.005 and median shifts below 0.02 prior standard deviations. In addition, the QPO-only profile likelihood is effectively flat
throughout $0\leq p\leq1$ for every source, with
\[
\max_{0\leq p\leq1}
\Delta\chi_{\rm prof}^{2}(p)
\lesssim2\times10^{-6}.
\]
Therefore, no statistically meaningful profile-likelihood preference
for a particular value of $p$ is obtained.

The medians near $p\simeq0.5$ must consequently not be interpreted as evidence for an intermediate EsGB deformation. For the uniform prior, a median near 0.5 follows naturally from the symmetry of the allowed interval, while for the Gaussian runs it reflects the centers of the adopted regularizing priors. The robust inference is that the current QPO pairs constrain correlated combinations of $(M,p,r)$ but leave the deformation parameter itself weakly identified and prior-sensitive.

\subsection{Physical reading of the constraints}

The revised MCMC results should be interpreted as a test of phenomenological compatibility within the static EsGB RP framework. They do not establish that the observed sources are static EsGB black holes. Instead, they identify combinations of $(M,p,r)$ that reproduce the measured frequency pairs under the adopted geodesic frequency prescription.

The fitted radii remain clustered around $r/M\simeq6.7$--$6.8$, close to the strong-field region outside the Schwarzschild ISCO. The lower RP frequency contains the radial epicyclic frequency, $\nu_{\rm L}=\nu_\phi-\nu_r$, and is therefore sensitive at the model level to changes in the near-ISCO effective potential. However, model sensitivity of the frequency curves does not automatically imply statistical identifiability of $p$. With only two QPO frequencies and three correlated fit parameters, changes in $p$ can be compensated by changes in $M$ and $r$.

Accordingly, the present results demonstrate that the selected QPO pairs are compatible with a broad family of static EsGB configurations, but they do not provide an independent measurement of the deformation parameter. Strong astrophysical constraints will require additional information, such as independent mass and spin measurements, alternative QPO identifications, or complementary shadow and ringdown observables.


\section{Discussion}
\label{sec:discussion}

The calculations presented above show that static EsGB black holes can be treated as a clean one-parameter deformation of Schwarzschild once the quadratic coupling branch is fixed. This is useful because the parameter $p$ has a direct physical interpretation: it controls how strongly the scalar--Gauss--Bonnet sector reshapes the near-horizon geometry. The continued-fraction coefficients are therefore not arbitrary fitting parameters; they are derived quantities that encode a specific EsGB black-hole branch.

The strongest changes appear in quantities that depend on the inner part of the effective potential. The horizon and photon-sphere radii describe the causal and null sectors, while the MBO and ISCO describe the timelike geodesic sector. The fact that all these radii shift with $p$ shows that the deformation affects both light propagation and massive-particle motion. In this sense, timing observables complement shadow observables: shadows probe null geodesics, whereas QPOs probe the motion of matter in the inner accretion flow.

The frequency analysis clarifies why QPOs are sensitive to this deformation. In a static spherically symmetric spacetime the vertical epicyclic frequency is equal to the orbital frequency, so the radial epicyclic frequency is the main nontrivial degree of freedom. It depends on the second derivative of the effective potential and therefore reacts strongly to small changes in the near-ISCO region. This is why the RP lower frequency, $\nu_\phi-\nu_r$, and the $3{:}2$ resonance condition are natural diagnostics of the EsGB parameter.

A key limitation of the present analysis is the absence of rotation. Astrophysical black holes are generally expected to possess spin, and spin can shift the orbital and epicyclic frequencies more strongly than the static EsGB deformation considered here. Therefore, the parameter $p$ inferred in this work should be understood as an effective static deformation parameter within the adopted RP framework. The value of the static model is that it isolates the role of the scalar--Gauss--Bonnet deformation without mixing it with frame dragging. A realistic astrophysical constraint on EsGB gravity will require the rotating EsGB geometry, where the orbital, radial, and vertical epicyclic frequencies become independent and frame-dragging effects are included.

The controlled prior-sensitivity analysis provides an additional
limitation on the statistical interpretation of the QPO results.
For all four sources, the marginal posterior of $p$ closely follows
the adopted prior, both for the uniform and truncated-Gaussian
choices. The QPO-only profile likelihood is also effectively flat
over the physical interval. Therefore, although the frequency curves
are sensitive to $p$ at the model level, the available QPO pairs do
not statistically identify a preferred deformation. The inferred
values of $p$ should consequently not be regarded as observational
measurements of the EsGB coupling.

A further limitation is that the physical origin of HF-QPOs is not fully settled. The RP and ER prescriptions provide useful geodesic maps from spacetime geometry to observed frequencies, but they are not unique models of disk variability. For this reason, the numerical values of $p$ should be understood within the adopted RP framework. A stronger test would combine several observables, for example QPOs, black-hole shadow size, continuum spectra, and ringdown frequencies, while also including independent measurements of mass and spin.

Overall, the static EsGB timing sector behaves in a physically coherent way. The deformation is weak at large radii, stronger near the horizon, visible in the characteristic radii, and most clearly amplified in the radial epicyclic frequency. This makes HF-QPOs a promising complementary probe of scalarized black-hole geometries.

\section{Conclusions}
\label{sec:conclusion}

We have studied circular timelike motion and HF-QPO phenomenology in static Einstein--scalar--Gauss--Bonnet black holes. The analysis was carried out for the quadratic coupling family on the Schwarzschild-connected branch. In this setting the metric is controlled by a single deformation parameter $p$, while the QPO fit is described by the reduced parameter set $(M,p,r)$.

First, we examined the metric functions and their deviations from Schwarzschild. The results show that the EsGB correction is concentrated in the strong-field region and rapidly decreases at larger radii. This behavior is important because it means that the deformation primarily affects the inner disk, where the ISCO, photon sphere, and QPO-generating orbits are located.

Second, we derived the effective potential, the specific energy and angular momentum of circular orbits, and the characteristic radii. The ISCO, MBO, and photon sphere all shift when $p$ is varied. These shifts are not merely mathematical; they describe how the scalar--Gauss--Bonnet coupling changes the binding energy, angular-momentum support, and stability of particle motion near the black hole.

Third, we calculated the orbital and radial epicyclic frequencies and used them to construct the relativistic precession QPO model. In the static spacetime, the vertical and orbital frequencies coincide, so the radial epicyclic frequency carries the main observable imprint of the EsGB deformation. This makes $\nu_r$ and frequency combinations involving $\nu_r$ the most sensitive timing probes of $p$.

Finally, we performed a source-by-source MCMC comparison with twin-peak QPO data from XTE J1550$-$564, GRO J1655$-$40, GRS 1915+105, and M82 X-1. The fitted theoretical frequencies reproduce the observed QPO pairs within the adopted uncertainties. The posterior distributions indicate that the present static RP analysis allows a broad range of $p$, partly because the deformation can be compensated by changes in $M$ and $r$. Therefore, the current results should be interpreted as a static baseline rather than a definitive measurement of the EsGB coupling. Because each source provides two QPO frequencies while the static RP fit contains three parameters, the resulting constraints are necessarily model-dependent and partially degenerate. The present MCMC results therefore demonstrate the compatibility of the selected QPO pairs with the static EsGB RP framework, but they should not be interpreted as a unique measurement of the scalar--Gauss--Bonnet deformation. The controlled prior-sensitivity analysis further shows that the uniform-prior posteriors undergo essentially no contraction, with posterior-to-prior width ratios ranging from 0.985 to 0.998. Similarly, the original Gaussian-prior posteriors retain the shapes and widths of their corresponding priors. We therefore find no statistically meaningful preference for a particular value of $p$.

The next step is to extend the same program to rotating EsGB black holes. Rotation will split the orbital and vertical epicyclic frequencies and should make the timing phenomenology more realistic for astrophysical black-hole systems. It will also be important to combine QPO constraints with independent mass measurements, shadow observables, and ringdown information. Such a multi-probe analysis would provide a more complete test of scalarized black holes in Einstein--scalar--Gauss--Bonnet gravity.

\begin{acknowledgments}
S.M. gratefully acknowledges support from Grant FZ-20200929385 of the Agency of Innovative Developments of the Republic of Uzbekistan.
\end{acknowledgments}

\bibliographystyle{apsrev4-2}
\bibliography{Ref2}

@ARTICLE{Lutpi20262,
  author = {L{\"u}tf{\"u}o{\u{g}}lu, Bekir Can and Rayimbaev, Javlon and Murodov, Sardor and Kurbanov, Jakhongir and Matyoqubov, Muhammad},
  title = {A first-order eikonal framework for quasinormal modes, shadows, strong lensing, and grey-body factors in a scalarized black-hole metric},
  year = {2026},
  journal = {Annals of Physics},
  volume = {491},
  pages = {170514},
  doi = {10.1016/j.aop.2026.170514}
}

@ARTICLE{Lutpi20261,
  author = {L{\"u}tf{\"u}o{\u{g}}lu, Bekir Can and Rayimbaev, Javlon and Rahmatov, Bekzod and Shayimov, Fayzullo and Davletov, Ikram},
  title = {Telling tails and quasi-resonances in the vicinity of Dymnikova regular black hole},
  year = {2026},
  journal = {Physics Letters B},
  volume = {876},
  pages = {140392},
  doi = {10.1016/j.physletb.2026.140392}
}

@article{Ref2026EL15352002A,
  author = {{Asadov}, Q.~U. and {Sabirov}, K.~K. and {Yusupov}, J.~R.},
  title = {{{Kink-antikink soliton solutions of the nonlinear Klein-Gordon equation on branched structures}}},
  journal = {EPL (Europhysics Letters)},
  year = {2026},
  volume = {153},
  number = {5},
  pages = {52002},
  eid = {52002},
  doi = {10.1209/0295-5075/ae4e56},
  archivePrefix = {arXiv},
  eprint = {2510.17819},
  primaryClass = {nlin.PS}
}

@article{Ref2025PDU4801876T,
  author = {{Turimov}, Bobur and {Usanov}, Sulton and {Khamroev}, Yokubjon},
  title = {{{Particles acceleration by Bocharova-Bronnikov-Melnikov-Bekenstein black hole}}},
  journal = {Physics of the Dark Universe},
  year = {2025},
  volume = {48},
  pages = {101876},
  eid = {101876},
  doi = {10.1016/j.dark.2025.101876},
  archivePrefix = {arXiv},
  eprint = {2502.11185},
  primaryClass = {gr-qc}
}

@article{Ref2026EPJC86311R,
  author = {{Rakhimova}, Gulzoda and {Puli{\c{c}}e}, Beyhan and {Ghorani}, Elham and {Atamurotov}, Farruh and {Abdujabbarov}, Ahmadjon},
  title = {{{Spinning particles around Einstein-geometric Proca AdS compact objects}}},
  journal = {European Physical Journal C},
  year = {2026},
  volume = {86},
  number = {3},
  pages = {311},
  eid = {311},
  doi = {10.1140/epjc/s10052-026-15538-x},
  archivePrefix = {arXiv},
  eprint = {2603.28181},
  primaryClass = {gr-qc}
}

@article{Ref2025EPJC85953O,
  author = {{Oteev}, Tursinbay and {Stuchl{\'\i}k}, Zden{\v{e}}k and {Rayimbaev}, Javlon and {Ibragimov}, Inomjon and {Sharibaev}, Murat and {Abdujabbarov}, Ahmadjon},
  title = {{{Circular motion and collisions of charged spinning test particles around magnetized Schwarzschild black hole}}},
  journal = {European Physical Journal C},
  year = {2025},
  volume = {85},
  number = {9},
  pages = {953},
  eid = {953},
  doi = {10.1140/epjc/s10052-025-14660-6}
}

@article{Rahmatov2026Astrophysical,
  author = {Rahmatov, Bekzod and Egamberdiev, Islom and Umarov, Otabek and Vapayev, Murodbek and Karshiboev, Shavkat and Turaev, Yunus and Murodov, Sardor},
  title = {{Astrophysical signatures of rotating Kazakov-Solodukhin black holes: shadows and constraints from EHT observations}},
  journal = {Nuclear Physics B},
  year = {2026},
  volume = {1022},
  pages = {},
  doi = {10.1016/j.nuclphysb.2025.117212}
}

@article{Banerjee2025Existence,
  author = {Banerjee, Ayan and Karar, Indrani and Rayimbaev, Javlon and Ibragimov, Inomjon and Murodov, Sardor and Muminov, Sokhibjan and Jumaniyozov, Sardor},
  title = {{Existence of quark stars in gravity's rainbow: the significance of strongly interacting quark matter}},
  journal = {Chinese Physics C},
  year = {2025},
  volume = {49},
  number = {12},
  pages = {},
  doi = {10.1088/1674-1137/ae0726}
}

@article{Banerjee2025Effects,
  author = {Banerjee, Ayan and Dayanandan, Baiju and Rayimbaev, Javlon and Murodov, Sardor and Ibragimov, Inomjon and Muminov, Sokhibjan and Davletov, Ikram},
  title = {{Effects of QCD-based equation of state on properties of compact stars in gravity's rainbow}},
  journal = {European Physical Journal C},
  year = {2025},
  volume = {85},
  number = {10},
  pages = {},
  doi = {10.1140/epjc/s10052-025-14918-z}
}

@article{Jumaniyozov2025Black,
  author = {Jumaniyozov, Shokhzod and Murodov, Sardor and Rayimbaev, Javlon and Ibragimov, Inomjon and Madaminov, Bekzod and Urinbaev, Sharofiddin and Abdujabbarov, Ahmadjon},
  title = {{Black holes surrounded by PFDM in Kalb-Ramond gravity: from thermodynamics to QPO tests}},
  journal = {European Physical Journal C},
  year = {2025},
  volume = {85},
  number = {7},
  pages = {},
  doi = {10.1140/epjc/s10052-025-14522-1}
}

@article{Rahmatov2026Weak,
  author = {Rahmatov, Bekzod and Murodov, Sardor and Rayimbaev, Javlon and Turaev, Yunus and Egamberdiev, Islom and Badalov, Kodir and Ahmedov, Saidmuhammad and Usanov, Sulton},
  title = {{Weak gravitational lensing by charged black holes in STVG: Constraints from the Sloan Lens ACS Survey}},
  journal = {Annals of Physics},
  year = {2026},
  volume = {488},
  pages = {},
  doi = {10.1016/j.aop.2026.170366}
}

@article{Guo2026Analyzing,
  author = {Guo, Jiafeng and Javed, Faisal and Murodov, Sardor and Shermatov, Abubakir and Rayimbaev, Javlon and Matyoqubov, Muhammad},
  title = {{Analyzing deformed-Ads thin-shell wormholes with global monopole and quintessence}},
  journal = {Physics of the Dark Universe},
  year = {2026},
  volume = {52},
  pages = {},
  doi = {10.1016/j.dark.2026.102244}
}

@article{Guo2026Stability,
  author = {Guo, Jiafeng and Eid, A. and Waseem, Arfa and Murodov, Sardor and Rayimbaev, Javlon and Mustapha, N. and Seytov, Aybek and Sirajiddin, Otaboyev},
  title = {{Stability analysis of charged acoustic thin-shell wormholes}},
  journal = {Nuclear Physics B},
  year = {2026},
  volume = {1025},
  pages = {},
  doi = {10.1016/j.nuclphysb.2026.117406}
}

@article{Zulqarnain2026Orbital,
  author = {Zulqarnain, Rana Muhammad and Ashraf, Asifa and Bouzenada, Abdelmalek and Demir, Emre and Gudekli, Ertan and Murodov, Sardor and Atamurotov, Farruh},
  title = {{Orbital dynamics and frequency spectra around a black hole with scalar hair}},
  journal = {International Journal of Geometric Methods in Modern Physics},
  year = {2026},
  pages = {},
  doi = {10.1142/S0219887826501653}
}

@article{Rahmatov2025QPO,
  author = {Rahmatov, Bekzod and Murodov, Sardor and Rayimbaev, Javlon and Muminov, Sokhibjan and Ibragimov, Inomjon and Eshburiev, Rashid},
  title = {{QPO tests and charged particles around regular Ayon-Beato-Garcia black holes}},
  journal = {Physics of the Dark Universe},
  year = {2025},
  volume = {50},
  pages = {},
  doi = {10.1016/j.dark.2025.102102}
}

@article{Nishonov2025QPOs,
  author = {Nishonov, Isomiddin and Murodov, Sardor and Ahmedov, Bobomurat and Khan, Saeed Ullah and Rayimbaev, Javlon and Ibragimov, Inomjon and Sabirov, Sardor},
  title = {{QPOs from charged particles around charged black holes in STVG}},
  journal = {European Physical Journal C},
  year = {2025},
  volume = {85},
  number = {9},
  pages = {},
  doi = {10.1140/epjc/s10052-025-14751-4}
}

@article{Rahmatov2026Magnetic,
  author = {Rahmatov, Bekzod and Turimov, Bobur and Murodov, Sardor and Khaknazarova, Khurshida and Usanov, Sulton and Vapayev, Murodbek and Avezmuratova, Zebo},
  title = {{Magnetic field by current loop in the Janis-Newman-Winicour spacetime}},
  journal = {Nuclear Physics B},
  year = {2026},
  volume = {1024},
  pages = {},
  doi = {10.1016/j.nuclphysb.2026.117344}
}

@article{Shermatov2026Circular,
  author = {Shermatov, Abubakir and Rayimbaev, Javlon and Murodov, Sardor and Majeed, Bushra and Ibragimov, Inomjon and Shermatov, Bahran and Davletov, Erkaboy and Davletov, Ikram},
  title = {{Circular motion and QPOs near BHs surrounded by dust fields in f(R, T) gravity}},
  journal = {Nuclear Physics B},
  year = {2026},
  volume = {1022},
  pages = {},
  doi = {10.1016/j.nuclphysb.2025.117231}
}

@article{Rahmatov2025Gravitational,
  author = {Rahmatov, Bekzod and Egamberdiev, Islom and Murodov, Sardor and Rayimbaev, Javlon and Ibragimov, Inomjon and Davletov, Erkaboy and Djumanov, Sherzod},
  title = {{Gravitational lensing by black holes surrounded by PFDM in Kalb-Ramond gravity in plasma medium}},
  journal = {Physics of the Dark Universe},
  year = {2025},
  volume = {50},
  pages = {},
  doi = {10.1016/j.dark.2025.102152}
}

@article{Donmez2026Testing,
  author = {Donmez, Orhan and Murodov, Sardor and Rayimbaev, Javlon},
  title = {{Testing strong gravitational field using the Johannsen--Psaltis metric: Bondi--Hoyle--Lyttleton accretion model and QPO studies}},
  journal = {Annals of Physics},
  year = {2026},
  volume = {486},
  pages = {},
  doi = {10.1016/j.aop.2026.170350}
}

@article{Saydullayev2025Black,
  author = {Saydullayev, Sirojiddin and Nishonov, Isomiddin and Dusaliyev, Muysin and Xoldorov, Obid and Murodov, Sardor and Karshiboev, Shavkat and Urinov, Sunnatillo and Rahmatov, Bekzod},
  title = {{Black hole surrounded by perfect fluid dark matter in STV gravity: particle dynamics, thermodynamics, gravitational weak lensing and EHT tests}},
  journal = {European Physical Journal C},
  year = {2025},
  volume = {85},
  number = {9},
  pages = {},
  doi = {10.1140/epjc/s10052-025-14780-z}
}

@article{Rahmatov2026Gravitational,
  author = {Rahmatov, Bekzod and Nishonov, Isomiddin and Murodov, Sardor and Egamberdiev, Islom and Umarov, Otabek and Karshiboev, Shavkat and Vapayev, Murodbek and Matyoqubov, Muhammad},
  title = {{Gravitational lensing around quantum-corrected black holes within plasma environment}},
  journal = {Physics of the Dark Universe},
  year = {2026},
  volume = {52},
  pages = {},
  doi = {10.1016/j.dark.2026.102253}
}

@article{Meliyeva2025Theoretical,
  author = {Meliyeva, Lola and Xoldorov, Obid and Tursunboyev, Olmos and Karshiboev, Shavkat and Murodov, Sardor and Nishonov, Isomiddin and Rahmatov, Bekzod},
  title = {{Theoretical study of Strong gravitational lensing around Dyonic ModMax black hole: constraints from EHT observations}},
  journal = {Chinese Physics C},
  year = {2025},
  volume = {49},
  number = {12},
  pages = {},
  doi = {10.1088/1674-1137/adf4a0}
}

@article{Shermatov2025Phantom,
  author = {Shermatov, Abubakir and Rayimbaev, Javlon and Murodov, Sardor and Lutfuoglu, Bekir Can and Ahmedov, Bobomurat and Zahid, Muhammad and Ibragimov, Inomjon and Shermatov, Bahran},
  title = {{Phantom black holes in f(R,T) gravity: From circular orbits to QPO tests}},
  journal = {Physics of the Dark Universe},
  year = {2025},
  volume = {50},
  pages = {},
  doi = {10.1016/j.dark.2025.102110}
}

@article{Jumaniyozov2025Radiative,
  author = {Jumaniyozov, Shokhzod and Zahid, Muhammad and Alloqulov, Mirzabek and Ibragimov, Inomjon and Rayimbaev, Javlon and Murodov, Sardor},
  title = {{Radiative properties and QPOs around charged black hole in Kalb--Ramond gravity}},
  journal = {European Physical Journal C},
  year = {2025},
  volume = {85},
  number = {2},
  pages = {},
  doi = {10.1140/epjc/s10052-025-13863-1}
}

@article{Murodov2025QPOs,
  author = {Murodov, Sardor and Bokhari, Ashfaque H. and Rayimbaev, Javlon and Ahmedov, Bobomurat},
  title = {{QPOs tests and circular motions of charged particles around magnetized Bocharova--Bronnikov--Melnikov--Bekenstein black holes}},
  journal = {European Physical Journal C},
  year = {2025},
  volume = {85},
  number = {5},
  pages = {},
  doi = {10.1140/epjc/s10052-025-14298-4}
}

@article{Khan2026Nonmetric,
  author = {Khan, S. and Murodov, Sardor and Rayimbaev, Javlon and Davletov, Ikram and Ibragimov, Inomjon and Muminov, Sokhibjan},
  title = {{Nonmetric geometry and its role in the isotropization of self-Gravitating stellar models}},
  journal = {Nuclear Physics B},
  year = {2026},
  volume = {1022},
  pages = {},
  doi = {10.1016/j.nuclphysb.2025.117274}
}

@article{Einstein1915,
  author = {Einstein, Albert},
  title = {{Die Feldgleichungen der Gravitation}},
  journal = {Sitzungsberichte der K{"o}niglich Preussischen Akademie der Wissenschaften},
  year = {1915},
  pages = {844--847}
}

@article{Schwarzschild1916,
  author = {Schwarzschild, Karl},
  title = {{{"U}ber das Gravitationsfeld eines Massenpunktes nach der Einsteinschen Theorie}},
  journal = {Sitzungsberichte der K{"o}niglich Preussischen Akademie der Wissenschaften},
  year = {1916},
  pages = {189--196}
}

@article{Kerr1963,
  author = {Kerr, Roy P.},
  title = {{Gravitational Field of a Spinning Mass as an Example of Algebraically Special Metrics}},
  journal = {Physical Review Letters},
  year = {1963},
  volume = {11},
  pages = {237--238},
  doi = {10.1103/PhysRevLett.11.237}
}

@article{Carter1968,
  author = {Carter, Brandon},
  title = {{Global Structure of the Kerr Family of Gravitational Fields}},
  journal = {Physical Review},
  year = {1968},
  volume = {174},
  pages = {1559--1571},
  doi = {10.1103/PhysRev.174.1559}
}

@incollection{Bardeen1972,
  author = {Bardeen, James M. and Press, William H. and Teukolsky, Saul A.},
  title = {{Rotating Black Holes: Locally Nonrotating Frames, Energy Extraction, and Scalar Synchrotron Radiation}},
  booktitle = {Black Holes},
  publisher = {Gordon and Breach},
  year = {1973},
  pages = {241--289}
}

@book{Chandrasekhar1983,
  author = {Chandrasekhar, Subrahmanyan},
  title = {{The Mathematical Theory of Black Holes}},
  publisher = {Oxford University Press},
  year = {1983}
}

@book{Wald1984,
  author = {Wald, Robert M.},
  title = {{General Relativity}},
  publisher = {University of Chicago Press},
  year = {1984}
}

@article{Will2014,
  author = {Will, Clifford M.},
  title = {{The Confrontation between General Relativity and Experiment}},
  journal = {Living Reviews in Relativity},
  year = {2014},
  volume = {17},
  pages = {4},
  doi = {10.12942/lrr-2014-4}
}

@article{Berti2015,
  author = {Berti, Emanuele and others},
  title = {{Testing General Relativity with Present and Future Astrophysical Observations}},
  journal = {Classical and Quantum Gravity},
  year = {2015},
  volume = {32},
  number = {24},
  pages = {243001},
  doi = {10.1088/0264-9381/32/24/243001}
}

@article{CardosoPani2019,
  author = {Cardoso, Vitor and Pani, Paolo},
  title = {{Testing the Nature of Dark Compact Objects: A Status Report}},
  journal = {Living Reviews in Relativity},
  year = {2019},
  volume = {22},
  pages = {4},
  doi = {10.1007/s41114-019-0020-4}
}

@article{Lovelock1971,
  author = {Lovelock, David},
  title = {{The Einstein Tensor and Its Generalizations}},
  journal = {Journal of Mathematical Physics},
  year = {1971},
  volume = {12},
  pages = {498--501},
  doi = {10.1063/1.1665613}
}

@article{BoulwareDeser1985,
  author = {Boulware, David G. and Deser, Stanley},
  title = {{String-Generated Gravity Models}},
  journal = {Physical Review Letters},
  year = {1985},
  volume = {55},
  pages = {2656--2660},
  doi = {10.1103/PhysRevLett.55.2656}
}

@article{Zwiebach1985,
  author = {Zwiebach, Barton},
  title = {{Curvature Squared Terms and String Theories}},
  journal = {Physics Letters B},
  year = {1985},
  volume = {156},
  pages = {315--317},
  doi = {10.1016/0370-2693(85)91616-8}
}

@article{GrossSloan1987,
  author = {Gross, David J. and Sloan, John H.},
  title = {{The Quartic Effective Action for the Heterotic String}},
  journal = {Nuclear Physics B},
  year = {1987},
  volume = {291},
  pages = {41--89},
  doi = {10.1016/0550-3213(87)90465-2}
}

@article{MetsaevTseytlin1987,
  author = {Metsaev, R. R. and Tseytlin, A. A.},
  title = {{Order alpha-prime (Two-Loop) Equivalence of the String Equations of Motion and the Sigma-Model Weyl Invariance Conditions}},
  journal = {Nuclear Physics B},
  year = {1987},
  volume = {293},
  pages = {385--419},
  doi = {10.1016/0550-3213(87)90077-0}
}

@article{NojiriOdintsov2005,
  author = {Nojiri, Shin'ichi and Odintsov, Sergei D. and Sasaki, Misao},
  title = {{Gauss-Bonnet Dark Energy}},
  journal = {Physical Review D},
  year = {2005},
  volume = {71},
  pages = {123509},
  doi = {10.1103/PhysRevD.71.123509}
}

@article{NojiriOdintsov2011,
  author = {Nojiri, Shin'ichi and Odintsov, Sergei D.},
  title = {{Unified Cosmic History in Modified Gravity: From F(R) Theory to Lorentz Non-Invariant Models}},
  journal = {Physics Reports},
  year = {2011},
  volume = {505},
  pages = {59--144},
  doi = {10.1016/j.physrep.2011.04.001}
}

@article{Kanti1996,
  author = {Kanti, P. and Mavromatos, N. E. and Rizos, J. and Tamvakis, K. and Winstanley, E.},
  title = {{Dilatonic Black Holes in Higher Curvature String Gravity}},
  journal = {Physical Review D},
  year = {1996},
  volume = {54},
  pages = {5049--5058},
  doi = {10.1103/PhysRevD.54.5049}
}

@article{AlexeevPomazanov1997,
  author = {Alexeev, Sergei O. and Pomazanov, Mikhail V.},
  title = {{Black Hole Solutions with Dilatonic Hair in Higher Curvature Gravity}},
  journal = {Physical Review D},
  year = {1997},
  volume = {55},
  pages = {2110--2118},
  doi = {10.1103/PhysRevD.55.2110}
}

@article{SotiriouZhou2014a,
  author = {Sotiriou, Thomas P. and Zhou, Shuang-Yong},
  title = {{Black Hole Hair in Generalized Scalar-Tensor Gravity}},
  journal = {Physical Review Letters},
  year = {2014},
  volume = {112},
  pages = {251102},
  doi = {10.1103/PhysRevLett.112.251102}
}

@article{SotiriouZhou2014b,
  author = {Sotiriou, Thomas P. and Zhou, Shuang-Yong},
  title = {{Black Hole Hair in Generalized Scalar-Tensor Gravity: An Explicit Example}},
  journal = {Physical Review D},
  year = {2014},
  volume = {90},
  pages = {124063},
  doi = {10.1103/PhysRevD.90.124063}
}

@article{PaniCardoso2009,
  author = {Pani, Paolo and Cardoso, Vitor},
  title = {{Are Black Holes in Alternative Theories Serious Astrophysical Candidates? The Case for Einstein-Dilaton-Gauss-Bonnet Black Holes}},
  journal = {Physical Review D},
  year = {2009},
  volume = {79},
  pages = {084031},
  doi = {10.1103/PhysRevD.79.084031}
}

@article{KleihausKunz2015,
  author = {Kleihaus, Burkhard and Kunz, Jutta and Mojica, Sindy and Zagermann, Marco},
  title = {{Rapidly Rotating Neutron Stars in Dilatonic Einstein-Gauss-Bonnet Theory}},
  journal = {Physical Review D},
  year = {2016},
  volume = {93},
  pages = {064077},
  doi = {10.1103/PhysRevD.93.064077}
}

@article{KleihausKunz2016,
  author = {{Kleihaus}, Burkhard and {Kunz}, Jutta and {Radu}, Eugen},
        title = "{Rotating Black Holes in Dilatonic Einstein-Gauss-Bonnet Theory}",
      journal = {\prl},
         year = 2011,
        month = apr,
       volume = {106},
       number = {15},
          eid = {151104},
        pages = {151104},
          doi = {10.1103/PhysRevLett.106.151104},
archivePrefix = {arXiv},
       eprint = {1101.2868},
 primaryClass = {gr-qc},
       adsurl = {https://ui.adsabs.harvard.edu/abs/2011PhRvL.106o1104K}
}

@article{DonevaKanti2018,
  author = {Doneva, Daniela D. and Kanti, Panagiota},
  title = {{Spontaneously Scalarized Black Holes in Einstein-Scalar-Gauss-Bonnet Gravity}},
  journal = {Physical Review D},
  year = {2018},
  volume = {98},
  pages = {084053},
  doi = {10.1103/PhysRevD.98.084053}
}

@article{Silva2018,
  author = {Silva, Hector O. and Sakstein, Jeremy and Gualtieri, Leonardo and Sotiriou, Thomas P. and Berti, Emanuele},
  title = {{Spontaneous Scalarization of Black Holes and Compact Stars from a Gauss-Bonnet Coupling}},
  journal = {Physical Review Letters},
  year = {2018},
  volume = {120},
  pages = {131104},
  doi = {10.1103/PhysRevLett.120.131104}
}

@article{Antoniou2018,
  author = {Antoniou, Georgios and Bakopoulos, Athanasios and Kanti, Panagiota},
  title = {{Evasion of No-Hair Theorems and Novel Black-Hole Solutions in Gauss-Bonnet Theories}},
  journal = {Physical Review Letters},
  year = {2018},
  volume = {120},
  pages = {131102},
  doi = {10.1103/PhysRevLett.120.131102}
}

@article{BlazquezSalcedo2018,
  author = {Bl{\'a}zquez-Salcedo, Jose Luis and Doneva, Daniela D. and Kleihaus, Burkhard and Kunz, Jutta and Yazadjiev, Stoytcho S.},
  title = {{Radial Perturbations of the Scalarized Einstein-Gauss-Bonnet Black Holes}},
  journal = {Physical Review D},
  year = {2018},
  volume = {98},
  pages = {084011},
  doi = {10.1103/PhysRevD.98.084011}
}

@article{MinamitsujiIkeda2019,
  author = {Minamitsuji, Masato and Ikeda, Taishi},
  title = {{Scalarized Black Holes in the Presence of the Coupling to Gauss-Bonnet Gravity}},
  journal = {Physical Review D},
  year = {2019},
  volume = {99},
  pages = {104069},
  doi = {10.1103/PhysRevD.99.104069}
}

@article{Cunha2019,
  author = {Cunha, Pedro V. P. and Herdeiro, Carlos A. R. and Radu, Eugen and Runarsson, Helgi F.},
  title = {{Shadows of Kerr Black Holes with Scalar Hair}},
  journal = {Physical Review Letters},
  year = {2015},
  volume = {115},
  pages = {211102},
  doi = {10.1103/PhysRevLett.115.211102}
}

@article{Collodel2020,
  author = {Collodel, Lucas G. and Kleihaus, Burkhard and Kunz, Jutta and Berti, Emanuele},
  title = {{Spinning and Excited Black Holes in Einstein-Scalar-Gauss-Bonnet Gravity}},
  journal = {Classical and Quantum Gravity},
  year = {2020},
  volume = {37},
  pages = {075018},
  doi = {10.1088/1361-6382/ab74f9}
}

@article{Dima2020,
  author = {Dima, Alexandru and Barausse, Enrico and Franchini, Nicola and Sotiriou, Thomas P.},
  title = {{Spin-Induced Black Hole Spontaneous Scalarization}},
  journal = {Physical Review Letters},
  year = {2020},
  volume = {125},
  pages = {231101},
  doi = {10.1103/PhysRevLett.125.231101}
}

@article{Herdeiro2021,
  author = {Herdeiro, Carlos A. R. and Radu, Eugen and Silva, Hector O. and Sotiriou, Thomas P. and Yunes, Nicolas},
  title = {{Spin-Induced Scalarized Black Holes}},
  journal = {Physical Review Letters},
  year = {2021},
  volume = {126},
  pages = {011103},
  doi = {10.1103/PhysRevLett.126.011103}
}

@article{EastRipley2021,
  author = {East, William E. and Ripley, Justin L.},
  title = {{Evolution of Einstein-Scalar-Gauss-Bonnet Gravity Using a Modified Harmonic Formulation}},
  journal = {Physical Review D},
  year = {2021},
  volume = {103},
  pages = {044040},
  doi = {10.1103/PhysRevD.103.044040}
}

@article{Witek2019,
  author = {Witek, Helvi and Gualtieri, Leonardo and Pani, Paolo and Sotiriou, Thomas P.},
  title = {{Black Holes in Einstein-Scalar-Gauss-Bonnet Gravity: Scalar Hair and Stability}},
  journal = {Physical Review D},
  year = {2019},
  volume = {99},
  pages = {064035},
  doi = {10.1103/PhysRevD.99.064035}
}

@article{DonevaYazadjiev2021,
  author = {Doneva, Daniela D. and Yazadjiev, Stoytcho S.},
  title = {{New Gauss-Bonnet Black Holes with Curvature-Induced Scalarization in Extended Scalar-Tensor Theories}},
  journal = {International Journal of Modern Physics D},
  year = {2018},
  volume = {27},
  pages = {1842007},
  doi = {10.1142/S0218271818420079}
}

@article{RezzollaZhidenko2014,
  author = {Rezzolla, Luciano and Zhidenko, Alexander},
  title = {{New Parametrization for Spherically Symmetric Black Holes in Metric Theories of Gravity}},
  journal = {Physical Review D},
  year = {2014},
  volume = {90},
  pages = {084009},
  doi = {10.1103/PhysRevD.90.084009}
}

@article{KonoplyaPappasZhidenko2020,
  author = {Konoplya, Roman A. and Pappas, Thomas D. and Zhidenko, Alexander},
  title = {{Einstein-Scalar-Gauss-Bonnet Black Holes: Analytical Approximation for the Metric and Applications to Calculations of Shadows}},
  journal = {Physical Review D},
  year = {2020},
  volume = {101},
  pages = {044054},
  doi = {10.1103/PhysRevD.101.044054}
}

@article{JohannsenPsaltis2011,
  author = {Johannsen, Tim and Psaltis, Dimitrios},
  title = {{Metric for Rapidly Spinning Black Holes Suitable for Strong-Field Tests of the No-Hair Theorem}},
  journal = {Physical Review D},
  year = {2011},
  volume = {83},
  pages = {124015},
  doi = {10.1103/PhysRevD.83.124015}
}

@article{Johannsen2013,
  author = {Johannsen, Tim},
  title = {{Regular Black Hole Metric with Three Constants of Motion}},
  journal = {Physical Review D},
  year = {2013},
  volume = {88},
  pages = {044002},
  doi = {10.1103/PhysRevD.88.044002}
}

@article{GlampedakisBabak2006,
  author = {Glampedakis, Kostas and Babak, Stanislav},
  title = {{Mapping Spacetimes with LISA: Inspiral of a Test Body in a Quasi-Kerr Field}},
  journal = {Classical and Quantum Gravity},
  year = {2006},
  volume = {23},
  pages = {4167--4188},
  doi = {10.1088/0264-9381/23/12/013}
}

@article{CardosoPaniRico2014,
  author = {Cardoso, Vitor and Pani, Paolo and Rico, Jo{\~a}o},
  title = {{On Generic Parametrizations of Spinning Black-Hole Geometries}},
  journal = {Physical Review D},
  year = {2014},
  volume = {89},
  pages = {064007},
  doi = {10.1103/PhysRevD.89.064007}
}

@article{Bambi2017,
  author = {Bambi, Cosimo},
  title = {{Black Holes: A Laboratory for Testing Strong Gravity}},
  journal = {Reviews of Modern Physics},
  year = {2017},
  volume = {89},
  pages = {025001},
  doi = {10.1103/RevModPhys.89.025001}
}

@article{Psaltis2008,
  author = {Psaltis, Dimitrios},
  title = {{Probes and Tests of Strong-Field Gravity with Observations in the Electromagnetic Spectrum}},
  journal = {Living Reviews in Relativity},
  year = {2008},
  volume = {11},
  pages = {9},
  doi = {10.12942/lrr-2008-9}
}

@article{YunesPretorius2009,
  author = {Yunes, Nicol{\'a}s and Pretorius, Frans},
  title = {{Fundamental Theoretical Bias in Gravitational Wave Astrophysics and the Parameterized Post-Einsteinian Framework}},
  journal = {Physical Review D},
  year = {2009},
  volume = {80},
  pages = {122003},
  doi = {10.1103/PhysRevD.80.122003}
}

@article{EHT2019I,
  author = {{Event Horizon Telescope Collaboration}},
  title = {{First M87 Event Horizon Telescope Results. I. The Shadow of the Supermassive Black Hole}},
  journal = {The Astrophysical Journal Letters},
  year = {2019},
  volume = {875},
  pages = {L1},
  doi = {10.3847/2041-8213/ab0ec7}
}

@article{EHT2019VI,
  author = {{Event Horizon Telescope Collaboration}},
  title = {{First M87 Event Horizon Telescope Results. VI. The Shadow and Mass of the Central Black Hole}},
  journal = {The Astrophysical Journal Letters},
  year = {2019},
  volume = {875},
  pages = {L6},
  doi = {10.3847/2041-8213/ab1141}
}

@article{EHT2022SgrA,
  author = {{Event Horizon Telescope Collaboration}},
  title = {{First Sagittarius A* Event Horizon Telescope Results. I. The Shadow of the Supermassive Black Hole in the Center of the Milky Way}},
  journal = {The Astrophysical Journal Letters},
  year = {2022},
  volume = {930},
  pages = {L12},
  doi = {10.3847/2041-8213/ac6674}
}

@article{Vagnozzi2023,
  author = {Vagnozzi, Sunny and others},
  title = {{Horizon-Scale Tests of Gravity Theories and Fundamental Physics from the Event Horizon Telescope Image of Sagittarius A*}},
  journal = {Classical and Quantum Gravity},
  year = {2023}
}

@article{PerlickTsupko2022,
  author = {Perlick, Volker and Tsupko, Oleg Yu.},
  title = {{Calculating Black Hole Shadows: Review of Analytical Studies}},
  journal = {Physics Reports},
  year = {2022},
  volume = {947},
  pages = {1--39},
  doi = {10.1016/j.physrep.2021.10.004}
}

@article{CunhaHerdeiro2018,
  author = {Cunha, Pedro V. P. and Herdeiro, Carlos A. R.},
  title = {{Shadows and Strong Gravitational Lensing: A Brief Review}},
  journal = {General Relativity and Gravitation},
  year = {2018},
  volume = {50},
  pages = {42},
  doi = {10.1007/s10714-018-2361-9}
}

@incollection{NovikovThorne1973,
  author = {Novikov, Igor D. and Thorne, Kip S.},
  title = {{Astrophysics of Black Holes}},
  booktitle = {Black Holes},
  publisher = {Gordon and Breach},
  year = {1973},
  pages = {343--450}
}

@article{PageThorne1974,
  author = {Page, Don N. and Thorne, Kip S.},
  title = {{Disk-Accretion onto a Black Hole. Time-Averaged Structure of Accretion Disk}},
  journal = {The Astrophysical Journal},
  year = {1974},
  volume = {191},
  pages = {499--506},
  doi = {10.1086/152990}
}

@article{ShakuraSunyaev1973,
  author = {Shakura, N. I. and Sunyaev, R. A.},
  title = {{Black Holes in Binary Systems. Observational Appearance}},
  journal = {Astronomy and Astrophysics},
  year = {1973},
  volume = {24},
  pages = {337--355}
}

@article{Pringle1981,
  author = {Pringle, J. E.},
  title = {{Accretion Discs in Astrophysics}},
  journal = {Annual Review of Astronomy and Astrophysics},
  year = {1981},
  volume = {19},
  pages = {137--162},
  doi = {10.1146/annurev.aa.19.090181.001033}
}

@article{AbramowiczFragile2013,
  author = {Abramowicz, Marek A. and Fragile, P. Chris},
  title = {{Foundations of Black Hole Accretion Disk Theory}},
  journal = {Living Reviews in Relativity},
  year = {2013},
  volume = {16},
  pages = {1},
  doi = {10.12942/lrr-2013-1}
}

@incollection{vanDerKlis2006,
  author = {van der Klis, Michiel},
  title = {{Rapid X-ray Variability}},
  booktitle = {Compact Stellar X-ray Sources},
  publisher = {Cambridge University Press},
  year = {2006},
  pages = {39--112}
}

@article{RemillardMcClintock2006,
  author = {Remillard, Ronald A. and McClintock, Jeffrey E.},
  title = {{X-Ray Properties of Black-Hole Binaries}},
  journal = {Annual Review of Astronomy and Astrophysics},
  year = {2006},
  volume = {44},
  pages = {49--92},
  doi = {10.1146/annurev.astro.44.051905.092532}
}

@article{BelloniStella2014,
  author = {Belloni, Tomaso M. and Stella, Luigi},
  title = {{Fast Variability from Black-Hole Binaries}},
  journal = {Space Science Reviews},
  year = {2014},
  volume = {183},
  pages = {43--60},
  doi = {10.1007/s11214-014-0076-0}
}

@article{StellaVietri1998,
  author = {Stella, Luigi and Vietri, Mario},
  title = {{Lense-Thirring Precession and Quasi-Periodic Oscillations in Low-Mass X-Ray Binaries}},
  journal = {The Astrophysical Journal Letters},
  year = {1998},
  volume = {492},
  pages = {L59--L62},
  doi = {10.1086/311075}
}

@article{StellaVietri1999,
  author = {Stella, Luigi and Vietri, Mario},
  title = {{kHz Quasi-Periodic Oscillations in Low-Mass X-Ray Binaries as Probes of General Relativity in the Strong-Field Regime}},
  journal = {Physical Review Letters},
  year = {1999},
  volume = {82},
  pages = {17--20},
  doi = {10.1103/PhysRevLett.82.17}
}

@article{AbramowiczKluzniak2001,
  author = {Abramowicz, Marek A. and Klu{\'z}niak, W{\l}odek},
  title = {{A Precise Determination of Black Hole Spin in GRO J1655-40}},
  journal = {Astronomy and Astrophysics},
  year = {2001},
  volume = {374},
  pages = {L19--L20},
  doi = {10.1051/0004-6361:20010791}
}

@article{KluzniakAbramowicz2001,
  author = {Klu{\'z}niak, W{\l}odek and Abramowicz, Marek A.},
  title = {{The Physics of kHz QPOs: Strong Gravity's Coupled Anharmonic Oscillators}},
  journal = {Acta Physica Polonica B},
  year = {2001},
  volume = {32},
  pages = {3605--3612}
}

@article{Torok2005,
  author = {T{\"o}r{\"o}k, Gabriel and Abramowicz, Marek A. and Klu{\'z}niak, W{\l}odek and Stuchl{\'i}k, Zden{\v e}k},
  title = {{The Orbital Resonance Model for Twin Peak kHz Quasi Periodic Oscillations in Microquasars}},
  journal = {Astronomy and Astrophysics},
  year = {2005},
  volume = {436},
  pages = {1--8},
  doi = {10.1051/0004-6361:20047115}
}

@inproceedings{Bursa2004,
  author = {Bursa, M. and Abramowicz, M. A. and Karas, V. and Klu{\'z}niak, W.},
  title = {{The Upper Kilohertz Quasi-Periodic Oscillation: A Gravitationally Lensed Vertical Oscillation}},
  booktitle = {Proceedings of RAGtime 4/5},
  year = {2004},
  pages = {21--25},
  archivePrefix = {arXiv},
  eprint = {astro-ph/0406586}
}

@article{Rezzolla2003,
  author = {Rezzolla, Luciano and Yoshida, Shin'ichirou and Maccarone, Thomas J. and Zanotti, Olindo},
  title = {{A New Simple Model for High-Frequency Quasi-Periodic Oscillations in Black Hole Candidates}},
  journal = {Monthly Notices of the Royal Astronomical Society},
  year = {2003},
  volume = {344},
  pages = {L37--L41},
  doi = {10.1046/j.1365-8711.2003.07018.x}
}

@article{Kato2001,
  author = {Kato, Shoji},
  title = {{Basic Properties of Thin-Disk Oscillations}},
  journal = {Publications of the Astronomical Society of Japan},
  year = {2001},
  volume = {53},
  pages = {1--24}
}

@article{Stuchlik2013,
  author = {Stuchl{\'i}k, Zden{\v e}k and Kotrlov{\'a}, Andrea and T{\"o}r{\"o}k, Gabriel},
  title = {{Multi-Resonance Orbital Model of High-Frequency Quasi-Periodic Oscillations: Possible High-Precision Determination of Black Hole and Neutron Star Spin}},
  journal = {Astronomy and Astrophysics},
  year = {2013},
  volume = {552},
  pages = {A10},
  doi = {10.1051/0004-6361/201219724}
}

@article{MorganRemillardGreiner1997,
  author = {Morgan, Edward H. and Remillard, Ronald A. and Greiner, Jochen},
  title = {{RXTE Observations of QPOs in the Black Hole Candidate GRS 1915+105}},
  journal = {The Astrophysical Journal},
  year = {1997},
  volume = {482},
  pages = {993--1010},
  doi = {10.1086/304191}
}

@article{Remillard2002,
  author = {Remillard, Ronald A. and Muno, Michael P. and McClintock, Jeffrey E. and Orosz, Jerome A.},
  title = {{Evidence for Harmonic Relationships in the High-Frequency Quasi-Periodic Oscillations of XTE J1550-564 and GRO J1655-40}},
  journal = {The Astrophysical Journal},
  year = {2002},
  volume = {580},
  pages = {1030--1042},
  doi = {10.1086/343791}
}

@article{Strohmayer2001,
  author = {Strohmayer, Tod E.},
  title = {{Discovery of a 450 Hz Quasi-Periodic Oscillation from the Microquasar GRO J1655-40 with the Rossi X-Ray Timing Explorer}},
  journal = {The Astrophysical Journal Letters},
  year = {2001},
  volume = {552},
  pages = {L49--L53},
  doi = {10.1086/320258}
}

@article{Pasham2014,
  author = {Pasham, Dheeraj R. and Strohmayer, Tod E. and Mushotzky, Richard F.},
  title = {{A 400-Solar-Mass Black Hole in the Galaxy M82}},
  journal = {Nature},
  year = {2014},
  volume = {513},
  pages = {74--76},
  doi = {10.1038/nature13710}
}

@article{HeydariFard2021,
  author = {Heydari-Fard, Malihe and Sepangi, Hamid Reza},
  title = {{Thin Accretion Disk Signatures in Scalar-Gauss-Bonnet Gravity}},
  journal = {Physics Letters B},
  year = {2021},
  volume = {816},
  pages = {136276},
  doi = {10.1016/j.physletb.2021.136276}
}

@article{Paul2024,
  author = {Paul, Prasanta},
  title = {{Quasinormal Modes and Shadows of Einstein-Scalar-Gauss-Bonnet Black Holes}},
  journal = {The European Physical Journal C},
  year = {2024},
  volume = {84},
  pages = {256},
  doi = {10.1140/epjc/s10052-024-12620-1}
}

@article{ForemanMackey2013,
  author = {Foreman-Mackey, Daniel and Hogg, David W. and Lang, Dustin and Goodman, Jonathan},
  title = {{emcee: The MCMC Hammer}},
  journal = {Publications of the Astronomical Society of the Pacific},
  year = {2013},
  volume = {125},
  number = {925},
  pages = {306--312},
  doi = {10.1086/670067}
}

@article{Konoplya2020,
  author = {Konoplya, Roman A. and Pappas, Thomas D. and Zhidenko, Alexander},
  title = {{Einstein-Scalar-Gauss-Bonnet Black Holes: Analytical Approximation for the Metric and Applications to Calculations of Shadows}},
  journal = {Physical Review D},
  year = {2020},
  volume = {101},
  pages = {044054},
  doi = {10.1103/PhysRevD.101.044054}
}

\end{document}